\documentclass[aps,prev,preprintnumbers,floatfix,nofootinbib]{revtex4}
\usepackage{epsf,epsfig,subfigure,graphicx,amsmath,amssymb}
\usepackage{color}
\usepackage{mathrsfs}
\usepackage{verbatim}
\usepackage{natbib}

\begin{document}


\vskip.5cm
\title
{Radion-higgs mixed state at the LHC with \\ KK contributions 
to production and decay}

\author{Hirohisa Kubota$^1$}
\author{Mihoko Nojiri$^{1,2}$} 
\affiliation{
KEK Theory Center$^1$, Tsukuba, Ibaraki 305-0801, Japan \\
The Graduate University for Advanced Studies (Sokendai), \\
Department of Particle and Nuclear Physics, Tsukuba, Ibaraki 305-0801, Japan \\
IPMU${}^{1,2}$, Tokyo University, Kashiwanoha 5-1-5, Kashiwa, Chiba 277-8583, Japan 
}

\vglue 0.3truecm

\begin{abstract}
In this paper we study the higgs($h$) -radion($r$)  system of the Randall-Sundum model where 
matter fermion and  gauge fields live in the  bulk while the higgs boson resides at 
the infrared (IR)  brane. We take into account the Kalza-Klein(KK) loop correction to 
the $hVV$ and $rVV$ coupling where $V=g, \gamma$. Inspired by the LHC data in 2011 where a hint of the higgs boson has been seen at 125 GeV, we fix one of the mass eigenstates of the higgs-radion mixed state at 125~GeV, and scan over all the other parameters which have not been excluded by the 2011 higgs search data.  We find the regions of parameter space where the scalar particle at 125~GeV  has signal rate  significantly larger or smaller  than that of the Standard Model(SM) higgs.   The signal ratio Br($ZZ$)/Br($\gamma \gamma$) also may
deviate from that of the SM higgs boson.
\end{abstract}
\maketitle

\section{introduction}
The Standard Model(SM) has been tested by a variety of experiments for a long time,
but so far there are no serious differences between the SM predictions and the experimental data.
In spite of this success, the model is not satisfactory.
The gauge hierarchy problem is one of the theoretical faults of the model, and
there are many attempts to extended the SM to solve this problem. 
We focus on the Randall-Sundrum(RS) model\cite{Randall:1999ee,Antoniadis:1998ig} in this paper.

The ADD(Arkani-Hamed, Dimopoulos, Dvali) model tries to resolve the gauge hierarchy problem by introducing several large extra dimensions\cite{ArkaniHamed:1998rs}.
But, there appears a new hierarchy problem between the compactification scale of the large extra dimensions and the electroweak scale.  
The RS model can resolve the gauge hierarchy problem by introducing one small warped extra dimension. 
Due to the cosmological constant in the five-dimensional Einstein equation, there is a warp factor $e^{-kr_c \pi}$ in the RS metric, where $k$ is the five-dimensional curvature scale, and $r_c$  is the five-dimensional compactification radius.
Boundaries of the fifth dimension are two 3-branes which are called the ``visible'' and ``hidden'' brane.
The mass scale of the visible brane is suppressed by the warp factor.
When $kr_c \pi \sim 35$, TeV scale in four dimensional effective theory is generated from the fundamental Planck scale 
as long as there are no large scales among the fundamental parameters.
When we consider a new five dimensional scalar with boundary interaction terms, the distance between the two 3-branes can be stabilized by the Goldberger and Wise stabilized mechanism\cite{Goldberger:1999uk}.
 
In the original RS model, all the SM particles reside on the visible brane except for the graviton.
The higher-dimensional operators of the four dimensional effective theory are typically suppressed by 1/TeV, and corrections to 
the Flavor Changing Neutral Current(FCNC) and the precision electroweak measurements are large.
However, if the SM fermions and gauge bosons except for the higgs boson propagate in the extra dimension, we resolve this problem\cite{Davoudiasl:1999tf,Pomarol:1999ad,Casagrande:2008hr,Bauer:2009cf,Gherghetta:2000qt,Agashe:2004ay,Huber:2000ie,Agashe:2004cp,Agashe:2003zs}.
In this extended model, the higgs boson has to localize on the visible 3-brane to resolve the gauge hierarchy problem.
The zero-mode of the five dimensional particles become the SM particles and the higher-modes become the KK particles with TeV scale mass. 
From the FCNC and electroweak precision tests, the lightest KK gluon mass has to be larger than  3 TeV\cite{Agashe:2003zs}. 

 Five dimensional fermion wavefunctions are determined by  the five dimensional fermion Dirac mass terms\cite{Gherghetta:2000qt,Chang:1999nh,Grossman:1999ra}.
If the fermion wavefunctions of the first and second generation KK particles peak near the hidden brane, the FCNC and the precision electroweak corrections can be suppressed by the Planck scale.
The hierarchy among the fermion masses is also relaxed in this model because the Yukawa couplings are product of a fifth dimensional Yukawa coupling and the values of wavefunctions at the visible brane.

Another interesting feature of the model is a gluon KK decay pattern.
Since the fifth dimensional wavefunction of the gluon and the top quark localize to the visible brane, the coupling between the lightest KK gluon $g_{kk}^{(1)}$ and the top quark is large. 
The KK gluon $g_{kk}^{(1)}$ dominantly decays into $t \bar{t}$\cite{Lillie:2007yh,Agashe:2006hk}.
Therefore, the process $pp \rightarrow g_{kk}^{(1)} \rightarrow t \bar{t}$ may be observed at the LHC. 
From the searches at the CMS, the lightest KK gluon mass has to be larger than 1.5 TeV\cite{Rappoccio:2011nj}.
This is the best limit from the direct KK particle production to the model.

In the four dimensional effective theory of the RS-type model, there is a scalar field which is called "radion" corresponding to fluctuations of the two 3-branes distance. 
The radion is massless before the stabilization of the branes.
The radion obtains  a finite mass due to the stabilization of the two branes and can be the lightest new particle in this model. It is possible to introduce a radion-higgs mixing term in the 
four dimensional effective action. 
Since interactions between the radion and the SM particles are similar to those of the higgs boson, it has been constrained by the higgs boson search at the LHC.
Sometimes, a
large deviation of the higgs signal rate from the SM prediction
may be observed\cite{Csaki:2000zn,Dominici:2002jv,Csaki:2007ns,Barger:2011qn,Grzadkowski:2012ng,deSandes:2011zs}.

There are contributions from KK states in the effective $r$($h$)-$V$-$V$ couplings 
though triangle loop diagrams.
The KK $W$ boson and the KK fermion towers contribute to these processes.
The effect in production and decay of the higgs boson is calculated in 
\cite{Casagrande:2008hr,Casagrande:2010si,Goertz:2011hj,Carena:2012fk}.
Couplings of the higgs boson with these KK particles may be large in some region of parameter space.  
The maximal effect to the higgs production nd decay is estimated in  \cite{Carena:2012fk}, and quite sizable. 
The correction could affect the phenomenology of the radion-higgs mixed system as well. 
In addition, radion-KK couplings should induce some effect to the radion-$gg$  and $\gamma \gamma$ couplings as well. 

In this paper, we revisit  the phenomenology  of  the radion-higgs mixed system where the SM gauge bosons and matter fermions are bulk fields.  
We include a moderate KK correction from the top sector in our  calculation
both for the higgs boson and  the radion effective couplings. There are two higgs-radion 
mixed scalars in this model.  In this paper,  
One of the scalar masses is fixed to 125~GeV in our calculation, using the 2011 data as the hint, while the other 
scalar mass and mixing are scanned.  There limits obtained from 2011 data significantly restrict the parameter 
region. Nevertheless, we find regions where deviation of the signal from the SM expectation is large enough 
so that it  would be observed by the expected luminosity in the 2012 LHC run.
 
The paper is organized as follows In section 2, and section 3 we explain the set up of our model, describe the  KK reduction of bulk fields,  
and obtain the effective interaction of the radion and the higgs bosons. We also discuss the size 
of the KK correction in section 3. In section 4, the interactions of  the higgs-radion mixing states 
are summarized. . 
Section 5 is devoted to numeral calculations.  We present our result both  for a) a higgs-like scalar whose mass is 125 GeV 
and for b) a radion-like scalar whose mass is 125~GeV.  For both cases, we find regions where the mixed scalar with mass $\sim$125~GeV 
  is detectable and the branching ratios into $ZZ$ and $\gamma \gamma$ significantly deviate from the SM prediction. 
Section 6 is devoted for conclusion.

\section{Randall-Sundrum model}
The RS model can resolve the gauge hierarchy problem of one $AdS$ extra dimension
 since this model replaces the gauge hierarchy by an exponential function of  the five dimensional curvature $k$ times the size of the fifth dimension $r_c\pi$ which is called the warp factor $e^{-kr_c\phi}$.
The background metric of the original RS model is given by
\begin{equation}
ds^2=e^{-2kr_c\phi}\eta_{\mu \nu}dx^\mu dx^\nu+r_c^2 d\phi^2,
\end{equation}	
The fifth dimension has the topology $S_1/Z_2$ and is parametrized by $\phi(0<\phi<\pi)$.
The boundaries $\phi=0,\pi$ are locations of the four dimensional hidden($\phi=0$) and visible($\phi =\pi$) branes.
The SM particles localize on the visible brane in the original RS model.
When we derive the four dimensional effective action by integrating out the extra dimension,
the fundamental mass parameter scale in the visible brane is reduced from the Planck scale to the electroweak scale.
For $kr_c\pi \sim 35$, $m_{\rm{pl}}e^{-kr_c\phi} \sim O(1)$ TeV.

The value of $kr_c\pi$ is arbitrary before the stabilization of the fifth dimension . 
There is a mechanism for stabilizing the size of the extra dimension(Goldberger and Wise mechanism\cite{Goldberger:1999uk}).
In this mechanism,  a massive five dimensional scalar field with the interaction terms at the branes
generates the potential for the radion field in the effective theory. 
The value $kr_c\pi \sim 35$ can be achieved without large fine tuning among the parameters  of the higgs potential.

In the four dimensional effective theory,  a fluctuation of the fifth dimension radius become a scalar particle called the radion.
A radion having non-zero mass  can be lighter than other KK particles, because the 
the origin of its mass is different.
This particle may be observed at the LHC.
When we consider the scalar perturbation corresponding to the radion,
the five dimensional metric is written by
\begin{equation}
ds^2=e^{-2(kr_c\phi+F(x,\phi))}\eta_{\mu \nu}dx^\mu dx^\nu-(1+2F(x,\phi))^2r_c^2 d\phi^2,
\end{equation}	
where $F$ is scalar perturbation $F(x,\phi)=r(x)R(\phi)$\cite{Csaki:2000zn}.
Here, $r(x)$ is a four dimensional canonically normalized radion field and $R(\phi)$ is determined by Einstein equation.
When the back-reaction is small, $F(x,\phi)$ is given by
\begin{equation}
F(x,\phi)=\frac{r(x)}{\Lambda_\phi}e^{2kr_c(\phi-\pi)},
\end{equation}	 
where $\Lambda_\phi$ is the vacuum expectation value of the radion   $\Lambda_\phi=\sqrt{6}M_{pl}e^{-kr_c\pi} \sim$ TeV. 

In the original RS model, all fields except for the graviton localize on the four dimensional  visible 3-brane.
However, we may also allow the SM particles except the higgs boson to propagate in the fifth dimension.
There are advantages to the model because the effect of the higher dimensional operators are suppressed significantly.
When the SM fields propagate into the fifth dimension,
there are KK particles of bulk fields in the effective four dimensional theory. 

First, we illustrate the KK reduction of a  $U(1)$ bulk gauge boson\cite{Davoudiasl:1999tf,Pomarol:1999ad,Gherghetta:2000qt,Chang:1999nh}. 
Extending to a non-abelian gauge boson is trivial.
The five dimensional action of the bulk gauge boson is given by
\begin{align}
S_A=-\frac{1}{4}\int d^4x\int d\phi \sqrt{-G}G^{MK}G^{NL}F_{KL}F_{MN},
\end{align}	
where $G^{MN} (M,N=0,1,2,3,4)$ is the five dimensional metric and $\sqrt{-G}=e^{-4kr_c\phi}r_c$. 
We assume $A_\mu$ is $Z_2$ parity even and $A_4$ is $Z_2$ parity odd,
so that the action is $Z_2$ parity even.
The Z2 parity is assigned so that only the SM field remains for the zero modes. 
Because of this assignment, 
the $A_\mu$ have a zero mode in the four dimensional effective theory while the $A_4$ does not have zero mode because of the boundary condition.
Then, the five dimensional action of the gauge boson is described by
\begin{equation}
S_A=-\frac{1}{4}\int d^4x\int d\phi r_c [\eta^{\mu \kappa}\eta^{\nu \lambda}F_{\kappa \lambda}F_{\mu \nu}-2\eta^{\nu \lambda}A_\lambda \partial_4 (e^{-2kr_c\phi }\partial_4 A_\nu)].
\end{equation}

The KK reduction of $A_\mu$ is given by
\begin{equation}
A_\mu(x,\phi)=\sum_{n=0}A_\mu^{(n)}(x)\frac{\chi^{(n)}(\phi)}{\sqrt{r_c}},
\end{equation}
where $\chi^{(n)}$ is the fifth dimensional wavefunction of each KK level.

The four dimensional effective action is obtained 
by substituting $A_{\mu}$  the KK expansion Eq.(6) for the bulk action Eq.(5)
and integrateing over the fifth dimension.
The wavefunctions $\chi^{(n)}$ satisfy the following conditions;\\
$1)$ the orthonormality condition,
\begin{equation}
\int^\pi_{-\pi}d\phi \chi^{(m)}\chi^{(n)}=\delta^{mn}
\end{equation}
\\
$2)$ the bulk differential equation of each KK level,
\begin{equation}
-\frac{1}{r_c^2}\frac{d}{d\phi}\Big(e^{-2kr_c\phi}\frac{d}{d\phi} \chi^{(n)}\Big)=m_n^2 \chi^{(n)},
\end{equation}
where $m_n$ is  a mass of the n-th KK level.
The solution for $\chi^{(n)}$ is given by
\begin{align}
\chi^{(n)}&=\frac{e^{kr_c\phi}}{N_n}[J_1(z_n)+\alpha_n Y_1(z_n)], \\
\chi^{(0)}&=\frac{1}{\sqrt{2\pi}}, \\
\alpha_n&=-\frac{\pi}{2[\log(x_n/2)-kr_c\pi+\gamma+1/2]},
\end{align}
where $z_n=\frac{m_n}{k}e^{kr_c\phi}$, $x_n=\frac{m_n}{k}e^{kr_c\pi}$, and $\chi^{(0)}$ is the fifth dimensional zero mode($m_n=0$) wavefunction.
Moreover, the boundary conditions at the visible brane lead
\begin{equation}
J_1(x_n)+x_nJ_1^{\prime}(x_n)+\alpha_n[Y_1(x_n)+x_nY_1^{\prime}(x_n)]=0.
\end{equation}
The mass of each KK particle is determined by this equation.
For the KK reduction of the massive gauge bosons($W$,$Z$), there are small corrections
to the wavefunction and to the KK mass from the electroweak symmetry breaking.
In this paper, we neglect these corrections.

Next, we illustrate the KK reduction of the bulk massive fermions
\cite{Casagrande:2008hr,Bauer:2009cf,Gherghetta:2000qt}.
The action of the bulk fermions is given by
\begin{align}
S_f&=\int d^4x \int d\phi \sqrt{-G}\Big[e^M_a\Big\{\frac{i}{2}\bar{Q}\gamma^a\partial_MQ  \nonumber
-\partial_M\bar{Q}\gamma^aQ+\omega_{bcM}\bar{Q}\frac{1}{2}\{\gamma^a,\sigma^{bc}\}Q \\ \nonumber
&-m_{Q}sgn(\phi)\bar{Q}Q+\sum_{q=u,d}\Big(\frac{i}{2}\bar{q}\gamma^a\partial_Mq
-\partial_M\bar{q}\gamma^aq+\omega_{bcM}\bar{q}\frac{1}{2}\{\gamma^a,\sigma^{bc}\}q\Big)\\
& -\sum_{q=u,d}m_{q}sgn(\phi)\bar{q}q 
-\frac{v}{\sqrt{2}r_c}(\bar{u}_LY_{5D}u_R+\bar{d}_LY_{5D}d_R)\delta(|\phi|-\pi)\Big\} \Big],
\end{align}  
where $e^M_a=diag(e^{kr_c\phi},e^{kr_c\phi},e^{kr_c\phi},e^{kr_c\phi},1/r_c)$ is an inverse vielbein, $\omega_{bcM}$ is a spin connection, $u_{L,R}, d_{L,R}$ are five dimensional up and down type quarks, and $q$ is a $SU(2)_L$ singlet and $Q$ is a $SU(2)_L$ doublet of five dimensinal fields. 
The contribution of the spin connection vanishes. 
Since the higgs field is localized on the visible brane, the Yukawa coupling term in the last line of the Eq(13) is also localized on the visible brane. 
Each 4-dim spinor field in $Q$ has left and right components $Q_L, Q_R$.
The  left-handed components of $SU(2)_L$ doublet $Q$ must  $Z_2$ parity even while the 
right-handed components of  $SU(2)_L$ doublet are $Z_2$ parity odd.
Similarly,  the right-handed components of the singlet $q_{u,d}$ are $Z_2$ parity even though
the left-handed components of the singlet $q_{u,d}$ are $Z_2$ parity odd.
By integrating this action by parts, we obtain
\begin{align}
S_f&=\int d^4x \int d\phi r_c\Big[e^{-3kr_c\phi} \Big(\bar{Q}i\gamma_\mu \partial^\mu Q+\sum_  {q=u,d}\bar{q}i\gamma_\mu \partial^\mu q\Big) \\ \nonumber
&-e^{-4kr_c\phi} sgn(\phi)\Big(c_Qk\bar{Q}Q+\sum_{q=u,d}c_qk \bar{q}q\Big) \\ \nonumber
&-\frac{1}{2r_c}\Big\{\bar{Q}_L \Big(e^{-4kr_c\phi}\partial_\phi+\partial_\phi e^{-4kr_c\phi}\Big)Q_R
-\bar{Q}_R \Big(e^{-4kr_c\phi}\partial_\phi+\partial_\phi e^{-4kr_c\phi}\Big)Q_L \\
&+\sum_{q=u,d}\Big(\bar{q}_L \Big(e^{-4kr_c\phi}\partial_\phi+\partial_\phi e^{-4kr_c\phi}\Big)q_R \nonumber
-\bar{q}_R \Big(e^{-4kr_c\phi}\partial_\phi+\partial_\phi e^{-4kr_c\phi}\Big)q_L\Big)\Big\} \\ 
&-e^{-3kr_c\phi}\frac{v}{\sqrt{2}r_c}(\bar{u}_LY_{5D}u_R+\bar{d}_LY_{5D}d_R)\delta(|\phi|-\pi )\Big] \nonumber
\end{align}
Here $m_{Q,q}=c_{Q,q}k$.
We assume that left-handed and right-handed components have opposite $Z_2$ parity,
so that the action is even in the $Z_2$ parity. The bulk fermion field of even parity has the KK zero mode.
The KK reduction of the five dimensional fermion $\Psi$ is written by
\begin{equation}
\Psi(x,\phi)=\sum_n \psi_n^{L,R}(x)\frac{e^{2kr_c\phi}}{\sqrt{r_c}}\hat{f}_n^{L,R}(\phi),
\end{equation}   
where $\hat{f}_n^{L,(R)}(\phi)$ is the Left(Right)-handed fifth dimensional wavefunction of each KK level, and $\psi_n$ is the $n$-th order four dimensional fermion.
When we require the following conditions, we can obtain a canonical
four dimensional fermion action;\\
$1)$ The orthonormality condition of fermion
\begin{equation}
\int^\pi_{-\pi}e^{kr_c\phi}\hat{f}_n^{L*}(\phi)\hat{f}_n^L(\phi)
=\int^\pi_{-\pi}e^{kr_c\phi}\hat{f}_n^{R*}(\phi)\hat{f}_n^R(\phi)=\delta^{mn} \\
\end{equation}
\\
$2)$ The differential equation of $\hat{f}^{L,R}$ is
\begin{equation}
\Big(\pm \frac{1}{r_c}\partial_\phi-m\Big)\hat{f}_n^{L,R}(\phi)=-m_ne^{kr_c\phi}\hat{f}_n^{R,L}(\phi)+e^{kr_c\phi}\frac{v}{\sqrt{2}r_c}Y_{5D}\hat{f}_n^{R,L}\delta(\phi-\pi).
\end{equation}
As a gauge boson case, the fifth dimensional wavefunction of  the zero mode and the higher modes are given by
\begin{align}
\hat{f}_n^{L,R}(\phi)&=\frac{e^{kr_c\phi/2}}{N_n^{L,R}}\left(J_{c_{L,R}\mp 1/2}\right(\frac{m_n}{k}e^{kr_c\phi}\left)+\beta_n^{L,R}Y_{c_{L,R}\mp 1/2}\right(\frac{m_n}{k}e^{kr_c\phi}\left)\right), \\[6pt]
\hat{f}_0^{L,R}(\phi)&=\frac{e^{\pm c_{L,R}kr_c\phi}}{N_0^{L,R}}, \\[6pt]
N_0^{L,R}&=\sqrt{\frac{2[e^{kr_c\pi (1+2c_{L,R})}-1]}{kr_c(1+2c_{L,R})}}, \\[6pt]
(\beta^{L,R}_n)_{even}&=-\frac{(1/2+c_{L,R})J_{c_{L,R}\mp1/2}(m_n/k)+m_n/k \ J^{\prime}_{c_{L,R}\mp1/2}(m_n/k)}{(1/2+c_{L,R})Y_{c_{L,R}\mp1/2}(m_n/k)+m_n/k \ Y^{\prime}_{c_{L,R}\mp1/2}(m_n/k)}, \\[6pt]
(\beta^{L,R}_n)_{odd}&=-\frac{J_{c_{L,R}\mp1/2}(m_n/k)}{Y_{c_{L,R}\mp1/2}(m_n/k)},
\end{align}
where $(\beta^{L,R}_n)_{even}$ and $(\beta^{L,R}_n)_{odd}$ are determined by boundary conditions at $\phi=0$,
and we calculate $N_n^{L,R}$ numerically.
Here upper(lower) of $\pm$ or $\mp$ in equation (18)$\sim$(22) stands for $L(R)$.

When the SM fermions propagate in the  bulk, we can relax the hierarchy problem of the Yukawa couplings by the profile of the fifth dimensional fermion wavefunctions.
The profile is a function of the bulk fermion mass parameter.
When the absolute value of $c_{L,R}$ is small, the bulk fermion is localized to the visible brane. 
Since the higgs boson is localized on the visible brane, 
the fermion localized on the visible brane has a large coupling with the higgs boson. 
Conversely, a coupling between the higgs boson and a hidden brane-localized fermion is small.
When  the 3rd generation fermions localize to the visible brane and the 2nd and the 1st generation fermions localize to the hidden brane, the 3rd generation fermions have large Yukawa coupling and 1st and 2nd generation fermions have small Yukawa couplings.

\section{Interactions of radion and higgs boson with the SM fields and KK modes}
In this section, we illustrate the interactions of the radion and the higgs boson with the SM fields and the KK modes\cite{Csaki:2000zn,Dominici:2002jv,Csaki:2007ns,Barger:2011qn,Grzadkowski:2012ng,deSandes:2011zs}.
A mixing between the radion and the higgs boson will be discussed in Section 4.
The interactions of the radion with the SM fermions and the massive gauge bosons are very similar to that of the higgs boson but suppressed by $\Lambda_\phi$ instead of  the higgs vacuum expectation value $v=246$ GeV. However, there are additional tree and anomaly  contributions to the interaction with the massless gauge bosons. 
Then, the production cross section of the radion at the LHC is comparable to that of the higgs boson.

We take into account contributions of the KK states in the decay and production of the radion and the higgs.
These contributions appear in a triangle loop of decay processes $h,r \rightarrow gg,\gamma\gamma$ and the production process of the radion and the higgs boson $gg \rightarrow h,r$.
The KK states of fermions and $W$ boson contribute to these triangle loop processes.
For  the bulk SM particles, there are infinite towers of the higher KK modes.
Since the RS model is a cutoff theory,
we take into account the KK modes up to some finite higher-modes.

The radion and the higgs boson couple via the mass term of  the massive gauge bosons and fermions.
The interaction terms of the massive gauge bosons are given by
\begin{align}
\mathcal{L}_{r}^{WW,ZZ}&=-\frac{r(x)}{\Lambda_\phi}\Big[2M_W^2W_\mu^{(0)+}W^{(0)\mu -}+M_Z^2Z_\mu^{(0)}Z^{(0)} \\ \nonumber
&+4\pi M_W^2\sum_n\chi_W^{(n)}(\pi)\chi_W^{(n)}(\pi)W_\mu^{(n)+}(x)W^{(n)\mu-}(x)+...\Big], \\
 \mathcal{L}_{h}^{WW,ZZ}&=-\frac{h(x)}{v}\Big[2M_W^2W_\mu^{(0)+}W^{(0)\mu -}+M_Z^2Z_\mu^{(0)}Z^{(0)}\\
 &+4\pi M_W^2\sum_n\chi_W^{(n)}(\pi)\chi_W^{(n)}(\pi)W_\mu^{(n)+}(x)W^{(n)\mu-}(x)+... \Big],
 \nonumber
\end{align}
where $W_\mu^{(0)}(x)$ is the SM $W$ boson, $\chi_W^{(n)}(\pi)$ is a wavefunction of the $n$-th order KK mode on the visible brane($\phi=\pi$), and $W_\mu^{(n)}(x)$ is the $n$-th order KK $W$ boson.The first and second terms describe the SM interactions and the third term describes the interaction with the KK $W$ boson. We omit the charge neutral KK boson interactions, because they are irrelevant to our discussions.   
Since the KK gauge bosons localize near the visible brane, couplings of the higgs boson and the radion with the KK $W$ boson are large.  $\chi_W^{(n)}(\pi)\chi_W^{(n)}(\pi) \sim 3$ for all KK modes at $\Lambda_\phi=3$ TeV.

The interaction terms of a fermion are given by
\begin{align}
\mathcal{L}_{r}^{ff}&=\frac{r(x)}{\Lambda_\phi}\Big[m_f\bar{\psi}^{(0)}\psi^{(0)}
+e^{kr_c\pi}\frac{v}{\sqrt{2}r_c}Y_{5D}\sum_n\hat{f}^{(n)*}(\pi)\hat{f}^{(n)}(\pi)\bar{\psi}^{(n)}\psi^{(n)}\Big], \\
\mathcal{L}_{h}^{ff}&=\frac{h(x)}{v}\Big[m_f\bar{\psi}^{(0)}\psi^{(0)}
+e^{kr_c\pi}\frac{v}{\sqrt{2}r_c}Y_{5D}\sum_n\hat{f}^{(n)*}(\pi)\hat{f}^{(n)}(\pi)\bar{\psi}^{(n)}\psi^{(n)}\Big], 
\end{align}
where $\psi^{(0)}$ is the SM fermion field, $\hat{f}^{(n)}(\pi)$ is a wavefunction of the $n$-th order KK mode on the visible brane($\phi=\pi$), $\psi^{(n)}$ is the $n$-th order KK fermion, and $m_f=e^{kr_c\pi}\frac{v}{\sqrt{2}r_c}Y_{5D}\bar{\hat{f}}^{(0)}(\pi)\hat{f}^{(0)}(\pi)$.
Since, in our model,  the fermion mass hierarchy is created by the wavefunction,
the five dimensional Yukawa coupling is order one. 

Next, we illustrate couplings between the radion and the higgs boson to the SM massless gauge bosons($\gamma,g$).
Since the higgs boson can not couple to the massless gauge bosons in a tree level,
a effective coupling arising from  triangle diagram involving  fermions and the $W$ boson(in case of $\gamma \gamma$) is the lowest 
order. 
We call these contributions $\mathcal{L}_{\rm{triangle}}$. 
The interactions between the radion and the massless gauge bosons also have these contributions.
In case of $r,h \rightarrow \gamma \gamma$, the KK $W$ boson and the KK fermions contribute to the triangle loop.
In case of $r,h\rightarrow gg $ case, only the KK fermions contribute to the triangle loop.
The interactions of the KK modes with the radion and the the higgs boson in these processes are the third term of Eqs.(23) and (24) and the second term of Eqs.(25) and (26).
There are 
interactions between the KK mode and gauge boson $W_{KK}W_{KK} \rightarrow \gamma $, $f_{KK}f_{KK}\rightarrow \gamma, g $ 
in the triangle loop. These interactions
are the same to the SM gauge couplings due to the gauge invariance. These can be derived by using the orthonormality conditions Eqs.(7) and (16).

The radion has additional interaction to the gauge boson.
When the massless gauge bosons are allowed to propagate in the extra dimensions, the radion couple to the zero mode of the massless gauge boson in the tree level\cite{Csaki:2007ns}.
We call this contribution $\mathcal{L}_{\rm{tree}}$.
Moreover,  in the one loop level, there are also new contribution, which is proportional to beta function[19].
We call this contribution $\mathcal{L}_{\rm{anom}}$. 
The interaction terms of the radion with the SM massless gauge bosons
in non-mixing case
is given by
\begin{align}
\mathcal{L}_{\rm{tree}}&=-\frac{r(x)}{4\Lambda_\phi kr_c\pi}(F^a_{\mu \nu}F^{a\mu \nu}+F_{\mu \nu}F^{\mu \nu}), \\[6pt]
\mathcal{L}_{\rm{anom}}+\mathcal{L}_{\rm{triangle}}&=-\frac{r(x)}{4\Lambda_\phi}\Big(\frac{b_{QCD}^{r}\alpha_s}{2\pi}F^a_{\mu \nu}F^{a\mu \nu}+\frac{b_{EM}^{r}\alpha}{2\pi}F_{\mu \nu}F^{\mu \nu}\Big), \\[6pt]
\mathcal{L}_{r}^{\gamma \gamma,gg}&=\mathcal{L}_{\rm{triangle}}+\mathcal{L}_{\rm{tree}}+\mathcal{L}_{\rm{anom}} \\[6pt]
&=-\frac{r(x)}{4\Lambda_\phi}\Big[\Big(\frac{1}{kr_c\pi}+\frac{\alpha_s}{2\pi}b_{QCD}^{r}\Big)F^a_{\mu \nu}F^{a\mu \nu}+\Big(\frac{1}{kr_c\pi}+\frac{\alpha}{2\pi}b_{EM}^{r}\Big)F_{\mu \nu}F^{\mu \nu}\Big],
\end{align}
where
\begin{align}
b_{QCD}^{r}&=7+(F_f+F_f^{KK}),\\[6pt]
b_{EM}^{r}&=-11/3+8/3(F_f+F_f^{KK})-(F_W+F_W^{KK}), \\[6pt]
F_f&=\tau_f(1+(1-\tau_f)f(\tau_f)) \\[6pt]
F_W&=2+3\tau_W+3\tau_W(2-\tau_W)f(\tau_W), \\[6pt]
f(\tau)&=(Arc\sin\frac{1}{\sqrt{\tau}})^2  \ \rm{for} \ \tau > 1, \\[6pt]
f(\tau)&=-\frac{1}{4}(\log\frac{\eta_+}{\eta_-}-i\pi)^2 \ \rm{for} \ \tau < 1,  
\end{align}
and $\tau_i=\Big(\frac{2m_i}{m_r} \Big)^2$, $\eta_{\pm}=1\pm \sqrt{1-\tau} $.
$m_i$ is the mass of the particle in the triangle loop.

The sum of contributions of the SM fermion and W boson KK modes in the triangle loop are expressed by $F_f^{KK}$ and $F_W^{KK}$ respectively.
In case of the higgs, there is no term corresponding to $\mathcal{L}_{\rm{tree}}$ and $\mathcal{L}_{\rm{anom}}$.
The interaction terms between the higgs boson and the massless gauge bosons are given by
\begin{equation}
\mathcal{L}_{h}^{\gamma \gamma,gg}=-\frac{h(x)}{4v}\Big[\frac{\alpha_s}{2\pi}b_{QCD}^{h}F^a_{\mu \nu}F^{a\mu \nu}+\frac{\alpha}{2\pi}b_{EM}^{h}F_{\mu \nu}F^{\mu \nu}\Big],
\end{equation}
where $b_{QCD}^{h}=F_f+F_f^{KK}$, and $b_{EM}^{h}=8/3(F_f+F_f^{KK})-(F_W+F_W^{KK})$.

\begin{figure}[htbp]
 \begin{minipage}{0.45\hsize}
  \begin{center}
   \includegraphics[width=65mm]{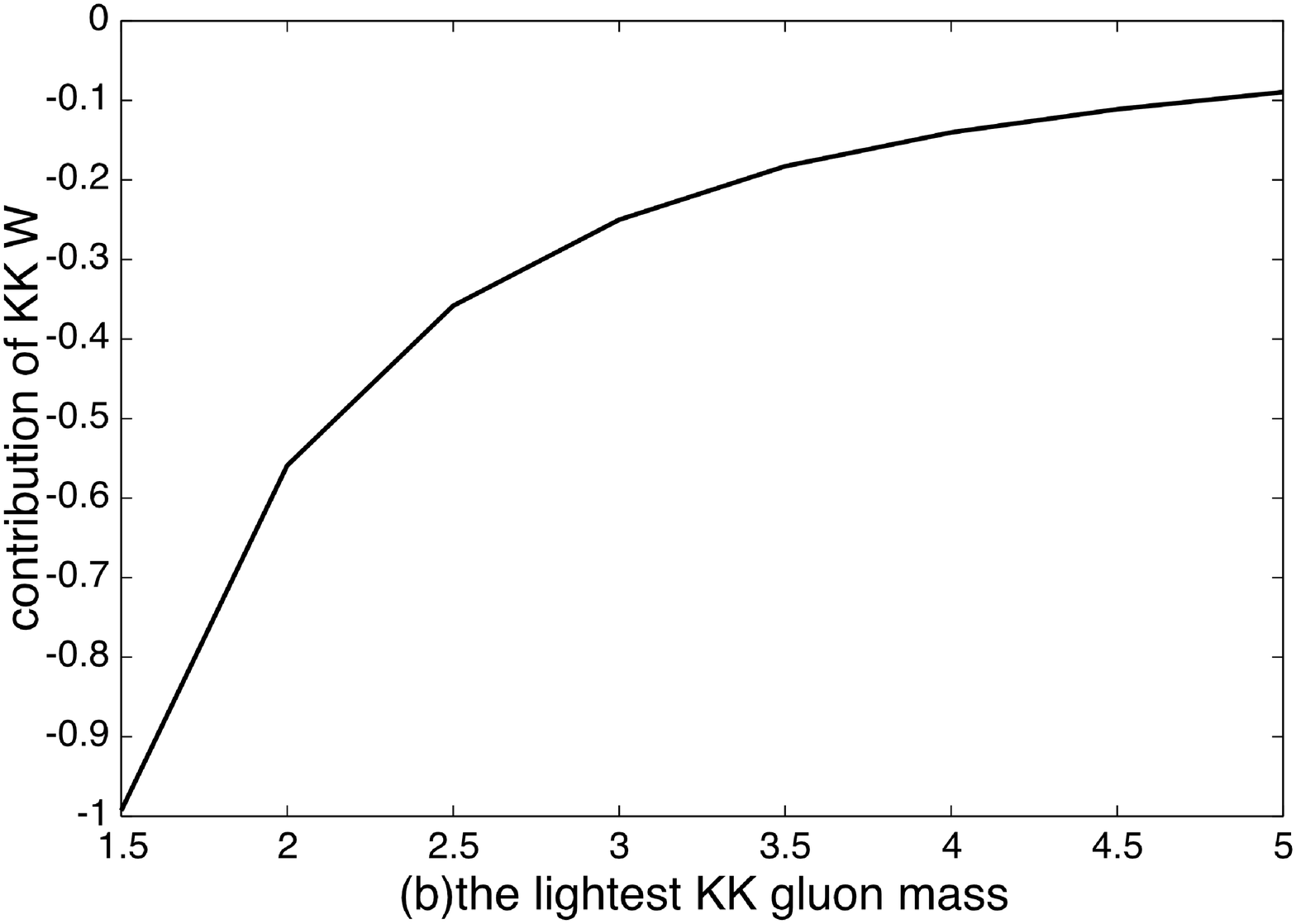}
  \end{center}
 \end{minipage}
 \begin{minipage}{0.45\hsize}
  \begin{center}
   \includegraphics[width=65mm]{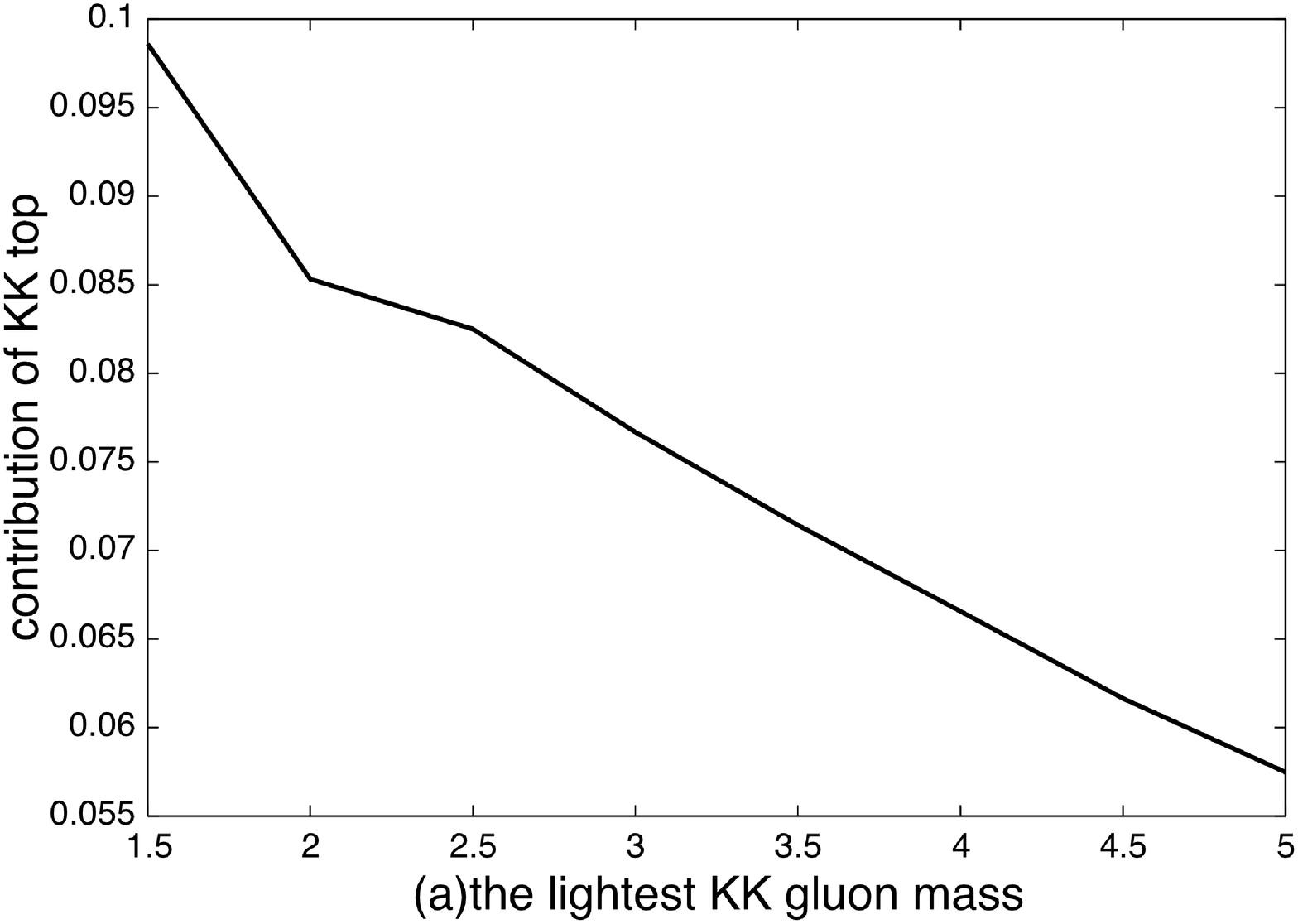}
  \end{center}
 \end{minipage}
 \caption{(a)$F_W^{KK}$ as a function of $\Lambda_\phi$.  (b) $F_f^{KK}$as a function of 
 $\Lambda_\phi$. only top KK contribution is included.
}\label{fig:one}

\end{figure}

The quantities $F_f^{KK}$ and $F_W^{KK}$are described in terms of $F_f$ and $F_W$ as follows.
\begin{align}
F_f^{KK}&=e^{kr_c\pi}\frac{v}{\sqrt{2}r_c}Y_{5D}\sum_n\frac{\hat{f}^{(n)*}(\pi)\hat{f}^{(n)}(\pi)}{m_{f_{KK}}^{(n)}}F_f, \\
F_W^{KK}&=2\pi \Big(\frac{M_W}{m_{W_{KK}}^{(n)}}\Big)^2\sum_n\chi_W^{(n)}(\pi)\chi_W^{(n)}(\pi)F_W,
\end{align}
where $m_{f_{KK}}^{(n)}$ and $m_{W_{KK}}^{(n)}$ are masses of the $n$-th KK levels.
The sum of KK state is finite\cite{Carena:2012fk}.
$F_W^{KK}$ depend on cutoff scale $\Lambda_\phi$, and
 $F_f^{KK}$ depend on both cutoff scale $\Lambda_\phi$ and $c_{L,R}$. 
since there are KK fermions in triangle loop.
We only take into account the contribution of the KK top in $F_f^{KK}$ in the numerical calculations of this paper. 
In paper \cite{Casagrande:2008hr,Casagrande:2010si,Goertz:2011hj,Carena:2012fk} it is pointed out that the effect of the 1st and 2nd generation fermions could be sizable, and we discuss the size of the effect  of $h \rightarrow gg$ coupling correction later.
We fix the five dimensional mass parameter of top quark $c_{L}=-47/100, c_{R}=28/100$,
when couplings of the KK top with the higgs and the radion are the largest.
Since the RS model is a theory with a cutoff, we sum the KK mode up to some 
 finite higher mode $n\sim60$\cite{Carena:2012fk}.
In FIG. \ref{fig:one}, we plot $F_W^{KK}$ and  $F_f^{KK}$ as  a function of the lightest KK gluon mass.
The lightest KK gluon mass is approximately equal to $\Lambda_\phi$.
As expected, $F_W^{KK}$ and  $F_f^{KK}$ are suppressed at large cutoff scale.
In $b^h_{QCD}=F_f+F^{KK}_f$, the effect of $F_f^{KK}$ could be sizable because $b^h_{QCD}/F_f \sim 0.85$ at $m_{h}=125$GeV and $\Lambda_\phi=3$~TeV. 
On the other hand, the effect to the radion coupling is subdominant given the large coefficient of the gauge boson anomaly term.
The $\Gamma(h \rightarrow gg)$ is described by $b_{QCD}^h$ only.
In the limit of zero higgs-radion mixing $\sigma(h\rightarrow gg)/\sigma(h\rightarrow gg)_{SM}$ at $m_h=125$ GeV is 0.78 for $\Lambda_\phi=$3TeV and our choice of $c_L$ and $c_R$.  

The radion production cross section without the radion-higgs mixing is suppressed by the cutoff scale $\Lambda_\phi$, but the direct coupling and the radion-$gg$ anomaly term enhance the production cross section;
$\sigma(r\rightarrow gg)/\sigma(h\rightarrow gg)_{SM}$=0.91 at $m_r=125$ GeV and $\Lambda_\phi$=3 TeV for our choice of $c_L$ and $c_R$.

In the paper \cite{Carena:2012fk}, the maximal KK correction to Br($h\rightarrow gg$) is considered by searching in the bulk fermion masses parameter space of all fermions. 
The effect is parametraized by $y^2_{max}$.  Suppression of the production ratio Br$(h\rightarrow gg)$/Br$(h\rightarrow gg)_{SM}$ as low as 0.1 can be achieved
for $y_{\rm{max}}\sim$3. This corresponds to the $b^h_{QCD}/F_f$  as low as 0.3. The value of $F^{KK}_f$ is factor 5 bigger than our canonical values. 
The effect of KK contribution in $b^r_{QCD}$  cannot be so big due to the large factor of 7 in Eq(32). For our canonical value, $F_f^{KK}$ is only 1.5$\%$ of total $b^r_{QCD}$, however, it could be as large as 7$\%$. This corresponds to the  13$\%$ correction to the  radion 
production ratio.   

A similar correction factor also appears in $b^h_{EM}$ and $b^r_{EM}$ , where the correction to the $ggh$ and $ggr$ coupling 
is now proportional to the ${q_i}^2$ where the $q_i$ are the charges of the KK fermions.  To scan over the parameters of the 
radion-higgs mixed system, it is convenient to parametrize the  KK loop corrections by up and down quark KK corrections $F^{KK}_u$ and $F^{KK}_d$.  
The effect to the $\gamma \gamma$  branching ratio would not be significant, because $F_W$ is relatively large,  $F_W$= 8.2.

Finally, the radion can also interact with the higgs boson.
The interaction Lagrangian between the radion and the higgs boson is given by
\begin{equation}
\mathcal{L}_{\rm{radion}}^{hh}=\frac{r(x)}{\Lambda_\phi}(-\partial_\mu h\partial^\mu h+2m_h^2 h^2)
\end{equation}
and it is used to calculate $\Gamma(r \rightarrow hh)$

\section{The radion-higgs mixing}
In this section, we consider radion-higgs mixing\cite{Csaki:2000zn,Dominici:2002jv,Barger:2011qn,Grzadkowski:2012ng,deSandes:2011zs}. The mixing is induced by a curvature-higgs mixing term in the four dimensional effective action
\begin{equation}
\mathcal{L}_\xi=\sqrt{g_{\rm{ind}}}\xi R(g_{\rm{ind}})H^\dag H,
\end{equation}
where $\xi$ is a mixing parameter, $g_{\rm{ind}}$ is a four dimensional induced metric $e^{-2(kr_c\pi+\frac{r(x)}{\Lambda_\phi})}\eta_{\mu \nu}$, and $R(g_{\rm{ind}})$ is a four dimensional Ricci scalar of the induced metric.

We introduce new variables for convenience.
\begin{align}
\gamma&=\frac{v}{\Lambda_\phi}, \\
\Omega(r)&=e^{-\frac{\gamma}{v}r(x)}.
\end{align}
Using these variables, $R(g_{\rm{ind}})$ is given by
\begin{equation}
R(\Omega^2(r)\eta_{\mu \nu})=-6\Omega^{-2}(\partial^2\log\Omega+(\nabla \log\Omega)^2)),
\end{equation}
where  $\Omega(r)$ can be expand by r(x) as  $\Omega(r)$$=1-\gamma r(x)/v+\cdot \cdot \cdot$.
By using Eq.(44),
we substitute $R(\Omega^2(r)\eta_{\mu \nu})$ for $\mathcal{L}_\xi$. We replace
$H \rightarrow (v+H)/\sqrt{2}$, 
$\mathcal{L}_\xi$ at quadratic order in $r$ is given by
\begin{equation}
\mathcal{L}_\xi=6\xi \gamma h\partial^2r +3\xi \gamma^2(\partial r)^2.
\end{equation}
In the four dimensional effective action, the higgs boson and  radion quadratic order terms are given by
\begin{equation}
\mathcal{L}=-\frac{1}{2}h\partial^2h-\frac{1}{2}m_h^2h^2-\frac{1}{2}(1+6\xi \gamma^2)r\partial^2r
-\frac{1}{2}m_r^2r^2+6\xi \gamma h\partial^2r,
\end{equation}
where $m_r$ and $m_h$ are the radion and the higgs mass parameters, 
and they coincide to the radion and the higgs mass for $\xi=0$. 
To obtain   canonical kinetic terms, we redefine the radion and the higgs field $h^{\prime},r^{\prime}$ as
\begin{align}
h&=h^{\prime}+6\xi\gamma r^{\prime}/Z, \\[6pt]
r&=r^{\prime}/Z, \\[6pt]
Z^2&=1+6\xi \gamma^2(1-6\xi). 
\end{align}
Since $Z^2$ has to be positive, the mixing parameter $\xi$ is restricted,
\begin{equation}
\frac{1}{12}\left(1-\sqrt{1+\frac{4}{\gamma^2}}\right) <\xi<\frac{1}{12}\left(1+\sqrt{1+\frac{4}{\gamma^2}}\right).
\end{equation}
We define the mass eigenstate as $h_m$ and $r_m$  so that 
\begin{align}
h^{\prime}&=\cos \theta h_m+\sin \theta r_m, \\
r^{\prime}&=-\cos \theta r_m+\sin \theta h_m, \\
\tan2 \theta&=12\xi \gamma Z \frac{m_h^2}{m_r^2-m_h^2(Z^2-36\xi^2 \gamma^2)}.
\end{align}
Then, the relation between the $r$, $h$ and the mass eigenstate $h_m$, $r_m$ is given by
\begin{align}
h&=(\cos \theta-\frac{6\xi \gamma}{Z}\sin \theta)h_m+(\sin \theta+\frac{6\xi \gamma}{Z}\cos\theta) r_m
\equiv d h_m+cr_m, \\
r&=\cos\theta\frac{r_m}{Z}-\sin\theta \frac{h_m}{Z}
\equiv ar_m+bh_m. 
\end{align}
The mass eigenvalues of the radion and higgs mixed scalars are given by 
\begin{equation}
m_{\pm}^2=\frac{1}{Z^2}\Big(m_r^2 +\beta m_h^2\pm \sqrt{(m_r^2 +\beta m_h^2)^2-4Z^2 m_r^2 m_h^2}\Big),
\end{equation}
where $\beta=1+6\xi \gamma^2$. 
Conversely, we can describe $(m_h,m_r)$ by using $m_{\pm}$.
\begin{equation}
(\beta m_h^2,m_r^2)=\frac{Z^2}{2}\Big[m_+^2+m_-^2\pm \sqrt{(m_+^2+m_-^2)^2-\frac{4\beta m_+^2m_-^2}{Z^2}}\Big]
\end{equation}
The quantity under the square root of Eq.(57) needs to be positive,
therefore the parameters must satisfy
\begin{equation}
(m_+^2+m_-^2)^2-\frac{4\beta m_+^2m_-^2}{Z^2} > 0
\end{equation}

These interactions are obtained by using Eqs.(54) and (55) and the interaction of 
$r$ and $h$ in section 3.
The interactions with massive gauge bosons and KK modes are given by
\begin{align}
\mathcal{L}_{r_m(\xi \neq 0)}^{WW,ZZ}&=-(c+\gamma a)\frac{r_m(x)}{v}\Big[2M_W^2W_\mu^{(0)+}W^{(0)\mu -}+M_Z^2Z_\mu^{(0)}Z^{(0)} \\ \nonumber
&+4\pi M_W^2\sum_n\chi_W^{(n)}(\pi)\chi_W^{(n)}(\pi)W_\mu^{(n)+}(x)W^{(n)\mu-}(x)+... \Big], \\
 \mathcal{L}_{h_m(\xi \neq 0)}^{WW,ZZ}&=-(d+\gamma b)\frac{h_m(x)}{v}\Big[2M_W^2W_\mu^{(0)+}W^{(0)\mu -}+M_Z^2Z_\mu^{(0)}Z^{(0)}\\ \nonumber
 &+4\pi M_W^2\sum_n\chi_W^{(n)}(\pi)\chi_W^{(n)}(\pi)W_\mu^{(n)+}(x)W^{(n)\mu-}(x)+... \Big],
\end{align} 
where $\gamma$ is Eq.(42). 
Similarly, the interactions with fermions and KK modes are given by
\begin{align}
\mathcal{L}_{r_m(\xi \neq 0)}^{ff}&=(c+\gamma a)\frac{r_m(x)}{v}\Big[m_f\bar{\psi}^{(0)}\psi^{(0)}
+e^{kr_c\pi}\frac{v}{\sqrt{2}r_c}Y_{5D}\sum_n\hat{f}^{(n)*}(\pi)\hat{f}^{(n)}(\pi)\bar{\psi}^{(n)}\psi^{(n)}\Big], \\
\mathcal{L}_{h_m(\xi \neq 0)}^{ff}&=(d+\gamma b)\frac{h_m(x)}{v}\Big[m_f\bar{\psi}^{(0)}\psi^{(0)}
+e^{kr_c\pi}\frac{v}{\sqrt{2}r_c}Y_{5D}\sum_n\hat{f}^{(n)*}(\pi)\hat{f}^{(n)}(\pi)\bar{\psi}^{(n)}\psi^{(n)}\Big], 
\end{align}
In case of massless gauge bosons, the radion interaction forms are different from the higgs boson.
The interactions with massless gauge bosons and KK modes  are given by
\begin{align}
\mathcal{L}_{r_m(\xi \neq 0)}^{\gamma \gamma,gg}&=-\frac{r_m(x)}{4v}\Big[\Big \{ \gamma a(\frac{1}{kr_c\pi}+\frac{\alpha_s}{2\pi}b_{QCD}^{r}\Big)+c \frac{\alpha_s}{2\pi}b_{QCD}^{h} \Big \}F^a_{\mu \nu}F^{a\mu \nu} \\ \nonumber
&+ \{\gamma a(\frac{1}{kr_c\pi}+\frac{\alpha_s}{2\pi}b_{EM}^{r}\Big)+ c \frac{\alpha_s}{2\pi}b_{EM}^{h} \Big \}F_{\mu \nu}F^{\mu \nu}\Big] \\
\mathcal{L}_{h_m(\xi \neq 0)}^{\gamma \gamma,gg}&=-\frac{h_m(x)}{4v}\Big[\Big \{\gamma b(\frac{1}{kr_c\pi}+\frac{\alpha_s}{2\pi}b_{QCD}^{r}\Big)+d \frac{\alpha_s}{2\pi}b_{QCD}^{h} \Big \}F^a_{\mu \nu}F^{a\mu \nu} \\
&+ \{\gamma b(\frac{1}{kr_c\pi}+\frac{\alpha_s}{2\pi}b_{EM}^{r}\Big)+d \frac{\alpha_s}{2\pi}b_{EM}^{h} \Big \}F_{\mu \nu}F^{\mu \nu}\Big] \nonumber
\end{align}

\section{Production and decay of the radion-higgs mixing state}
In this section, we consider radion-higgs phenomenology at the LHC.
Production and decay kinematics of the radion and the higgs boson are similar,
therefore, we may  observe a radion-higgs mixed state at the LHC in the near future.
Indeed, given the large quantity of data from the higgs search at the
ATLAS and the CMS in 2011, the radion-higgs mixing scenario is constrained already. 
Since the LHC is a $pp$ collider, the higgs boson and the radion are dominantly produced by the gluon fusion, and
they further decay into $WW$, $ZZ$ and $\gamma \gamma$.
We define detection ratios $DR_{h_m}(X)$ and $DR_{r_m}(X)$ as follows
\begin{align}
DR_{h_m}(X)&\equiv \frac{\Gamma_{h_m} (gg)BR(h_m \rightarrow X)}{\Gamma_{h_{SM}}(gg)BR(h_{SM} \rightarrow X)}, \\
DR_{r_m}(X)&\equiv \frac{\Gamma_{r_m} (gg)BR(r_m \rightarrow X)}{\Gamma_{h_{SM}}(gg)BR(h_{SM} \rightarrow X)},
\end{align}
where $X$ is $WW$, $ZZ$, and $\gamma \gamma$. 
The eigenstates $h_m$ and $r_m$ are defined in Eq.(54) and (55).
These quantities are the ratios of the signal from  $h_m$ ($r_m$)
production from gluon fusion and decay to $X$ to that of SM Higgs
boson.  
We can determine the excluded region of $h_m$ and $r_m$ by comparing $DR_{h_m}(X)$ and $DR_{r_m}(X)$ with experimental limits on $\sigma (h)/\sigma (h_{SM})$ because current limits are dominantly from $gg \rightarrow h$ production. 
Searches for  $h\rightarrow WW, ZZ$ at the LHC, $h\rightarrow ZZ\rightarrow 4l, h\rightarrow WW\rightarrow ll\nu\nu$ and $h\rightarrow ZZ\rightarrow ll\nu\nu$, give the strongest constraints in different mass ranges. 
We also take into account constraints of theoretical parameter space from Eq.(50) and (58).

The recent LHC data of the higgs search suggests excess of $ZZ$ and $\gamma \gamma$ signal at a mass near 125 GeV.\footnote{After submitting this paper the 
discovery of a new scalar is reported from both ATLAS and CMS.{\bf need references} }
Therefore, we use it to reduce the number of free parameters of our model.
We consider two scenarios in this paper.
In the Scenario I,  a higgs-like mixed state $h_m$ has mass $m_{h_m}=125$~GeV. A radion-like mixed state $r_m$ should satisfy the current experimental limits so that we will have not already discovered a second new scalars. 
We scan over $m_{r_m}$ and $\sin\theta$ which are the implicit functions of $m_r$, $m_h$ and $\xi$.
Conversely, in the Scenario II, a radion-like mixed state $r_m$ has
a mass $m_{r_m}=125$~GeV and a  higgs-like mixed state $h_m$ satisfy the
experimental limits. We call $h_m$ ($r_m$) of the scenario I (II) as
a visible  higgs boson $h_{\rm vis}$, and $r_m$ ($h_m$) as a hidden
scalar  $h_{\rm hid}$ in this paper. 
In both cases, if the detection ratio of $h_{\rm hid}$ exceed the 95$\%$ CL upper limit of the ATLAS search,
we regard the parameter values are excluded.\footnote{The CMS data is consistent with the ATLAS data and we do not
  use it. }

\begin{figure}[htbp]
 \begin{minipage}{0.45\hsize}
  \begin{center}
   \includegraphics[width=80mm]{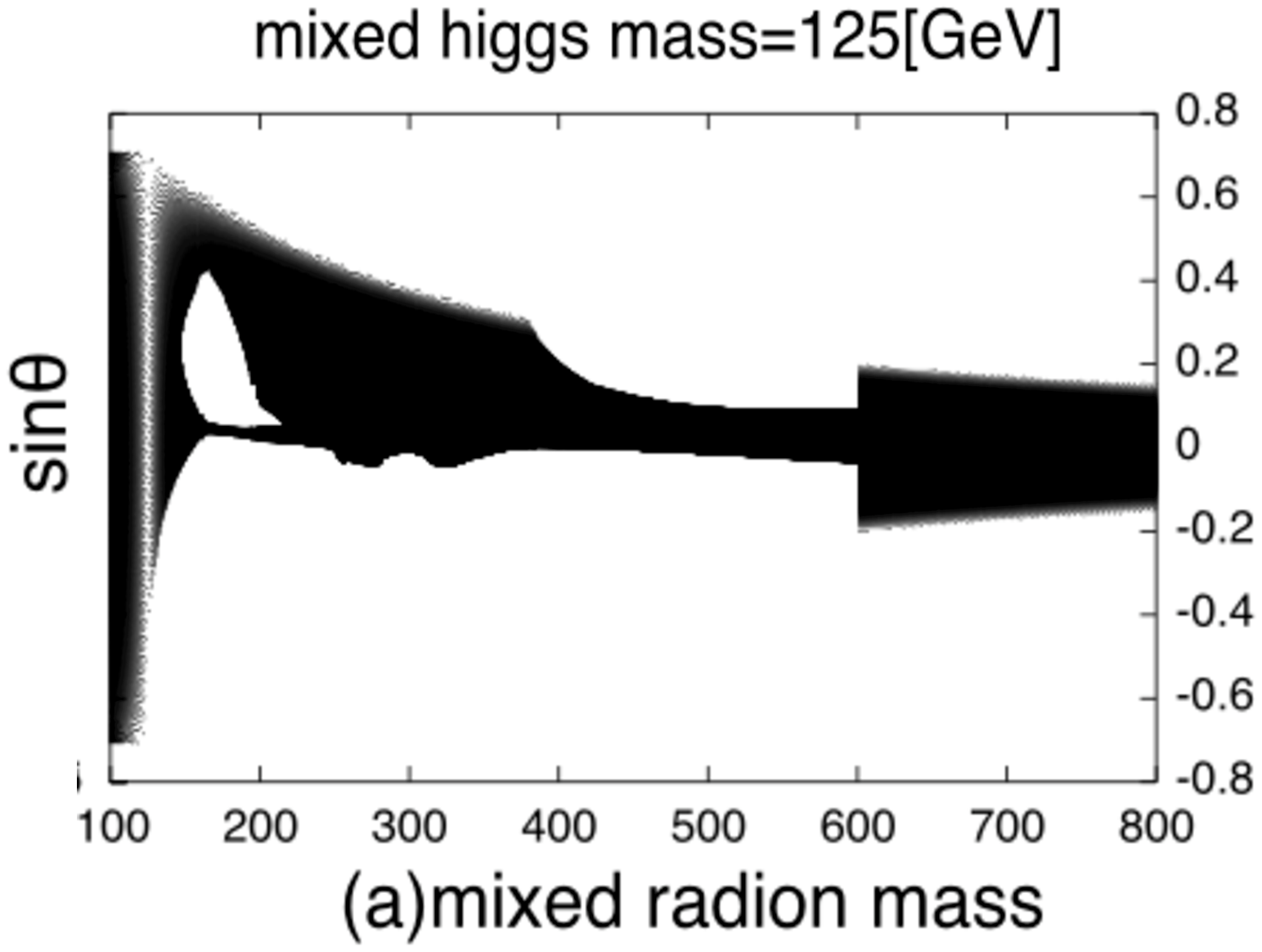}
  \end{center}
 \end{minipage}
 \begin{minipage}{0.45\hsize}
  \begin{center}
   \includegraphics[width=80mm]{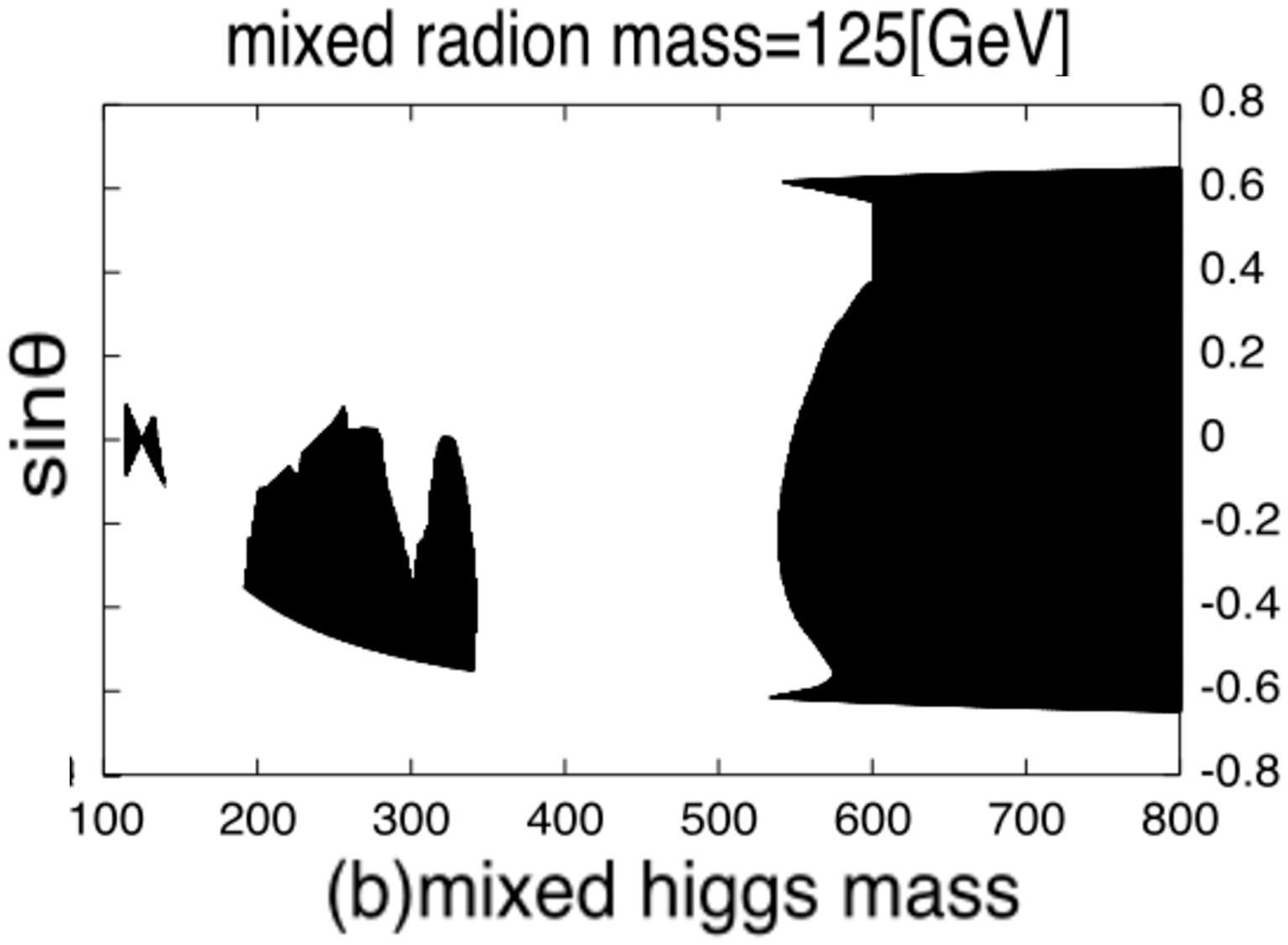}
  \end{center}
 \end{minipage}
\caption
{
The black regions shows allowed parameter region of (a) Scenario I  and
(b) Scenario II  as a function of the hidden scalar mass and mixing
angle($\sin\theta$).  We take $m_{h_{\rm vis}}=125$~GeV  and  $\Lambda_\phi=3$~TeV.
}
\label{fig:two}
\end{figure}

The parameter region which is not  excluded by the ATLAS data is shown  by black
regions in FIG. \ref{fig:two}(a) for Scenario I in the plane of hidden
scalar mass  $m_{r_m}$  and mixing angle($\sin\theta$) for
$\Lambda_\phi=3$~TeV and  $m_{h_m}$=125~GeV.
The mass of $r_m$ is not constrained in the figure, 
because the production cross section of the hidden scalar is suppressed by small mixing angle.
There is also the accidental cancelation of production cross section for small $m_{r_m}$ as we will see later. 

FIG. \ref{fig:two}(b)  shows the allowed region for Scenario II. Here
we fix $m_{r_m}=125$ GeV, and show the region which is not excluded in
$m_{h_m}$ and $\sin\theta$ plane.  
There are  also allowed regions for $m_{h_m}<130$~GeV, $180$~GeV$ <m_{h_m}<340$~GeV and $m_{h_m}>540$~GeV.

\begin{figure}[htbp]
 \begin{minipage}{0.5\hsize}
  \begin{center}
   \includegraphics[width=65mm]{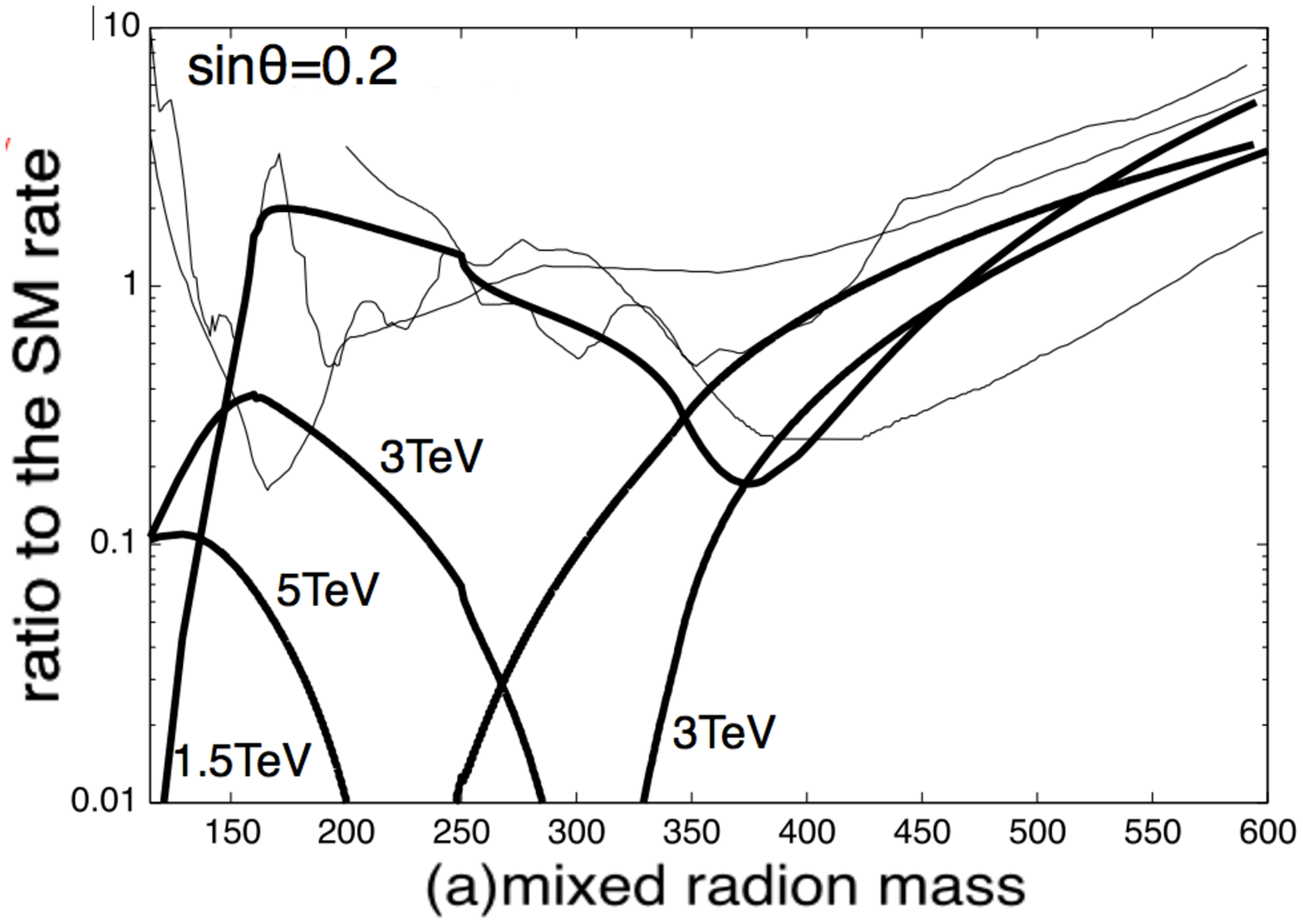}
  \end{center}
 \end{minipage}
 \begin{minipage}{0.45\hsize}
  \begin{center}
   \includegraphics[width=65mm]{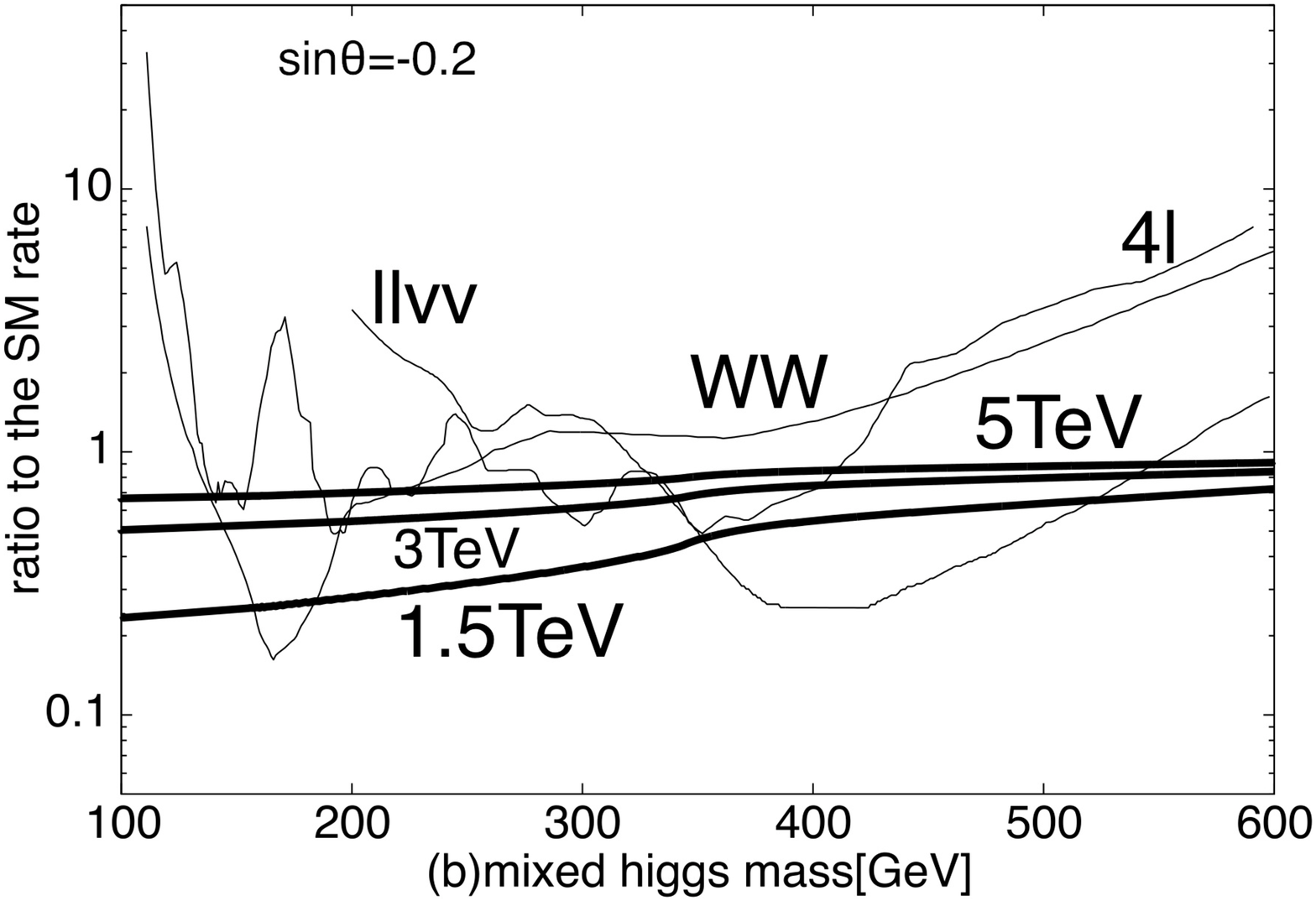}
  \end{center}
 \end{minipage}
 \caption
 {The detection ratio of hidden scalar as the function of its mass. 
 The thick lines of (a) shows $DR_{r_m}(WW,ZZ)$ for  $\Lambda_\phi=1.5, 3$ and 5~TeV.  The mixing angle is fixed to $\sin\theta=0.2$. Similarly, the thick lines of (b) are $DR_{h_m}(WW,ZZ)$ for $\Lambda_\phi=$1.5, 3,  and 5~TeV and $\sin\theta=-0.2$. The Thin lines are ATLAS 95$\%$ CL upper limit of $WW$,$ZZ$ $\rightarrow ll\nu\nu$ and $ZZ \rightarrow 4l$ channels. The visible higgs masses are fixed to be 125~GeV for both figures.  
 }
 \label{fig:three}
\end{figure}

To show the relation between the experimental constraints and detection ratios shown in FIG. \ref{fig:two},
we plot the detection ratio of a hidden scalar  FIG.\ref{fig:three} (a)$DR_{r_m}(WW,ZZ)$ for Scenario I 
and (b)$DR_{h_m}(WW,ZZ)$ for Scenario II as functions of the hidden scalar mass. The thick lines are the detection ratio for  $\Lambda_\phi=1.5$ up to 5 TeV. 
For both figure, the visible higgs mass is fixed to 
 125~GeV(for (a)$m_{h_m}$=125~GeV  and (b)$m_{r_m}$=125~GeV). 
 We take $\sin\theta=0.2(-0.2)$ 
 for FIG. \ref{fig:three} (a)  (FIG. \ref{fig:three} (b))  respectively.\footnote{In our model, $DR(WW)=DR(ZZ)$. }
The thin lines are ATLAS 95$\%$ upper limit of $WW$ and $ZZ$ channels.
In FIG. \ref{fig:three}(a) , $DR_{r_m}(WW, ZZ)$ is suppressed in the wide region.
The region with small $m_{r_m}$ remains  because of 
the accidental cancelation between $h \rightarrow gg$ and $r\rightarrow gg$ 
couplings for the mixed state.
When we change $m_{r_m}$ fixing $\theta$ and $m_{h_m}$, the parameter $\xi$ changes according to Eq.(55). The $\xi$ affects parameter $a,c$ appearing in the radion coupling in Eq.(61),(63) and (65) by Z defined in Eq.(49). 
FIG. \ref{fig:three}(b) show  the second scenario at $\sin\theta=-0.2$. 
$DR_{h_m}(WW, ZZ)$ is slightly suppressed.
The suppression mostly comes from the KK contributions to the  $h \rightarrow gg$ coupling
and we do not observe accidental cancelation.

Now we calculate the branching ratio of the visible higgs boson  $h_{\rm vis}$. 
We calculate $DR_{h_m}(ZZ)$ and $DR_{h_m}(\gamma \gamma)$ in the first scenario at $m_{h_m}=125$~GeV in all allowed parameter region with the excess of the higgs signal compared 
over  the SM expectation to the higgs boson.  
Similarly, in the second scenario, we calculate $DR_{r_m}(ZZ)$ and $DR_{r_m}(\gamma \gamma)$ at $m_{r_m}=125$ GeV.

\begin{figure}[htbp]
 \begin{minipage}{0.45\hsize}
  \begin{center}
   \includegraphics[width=68mm]{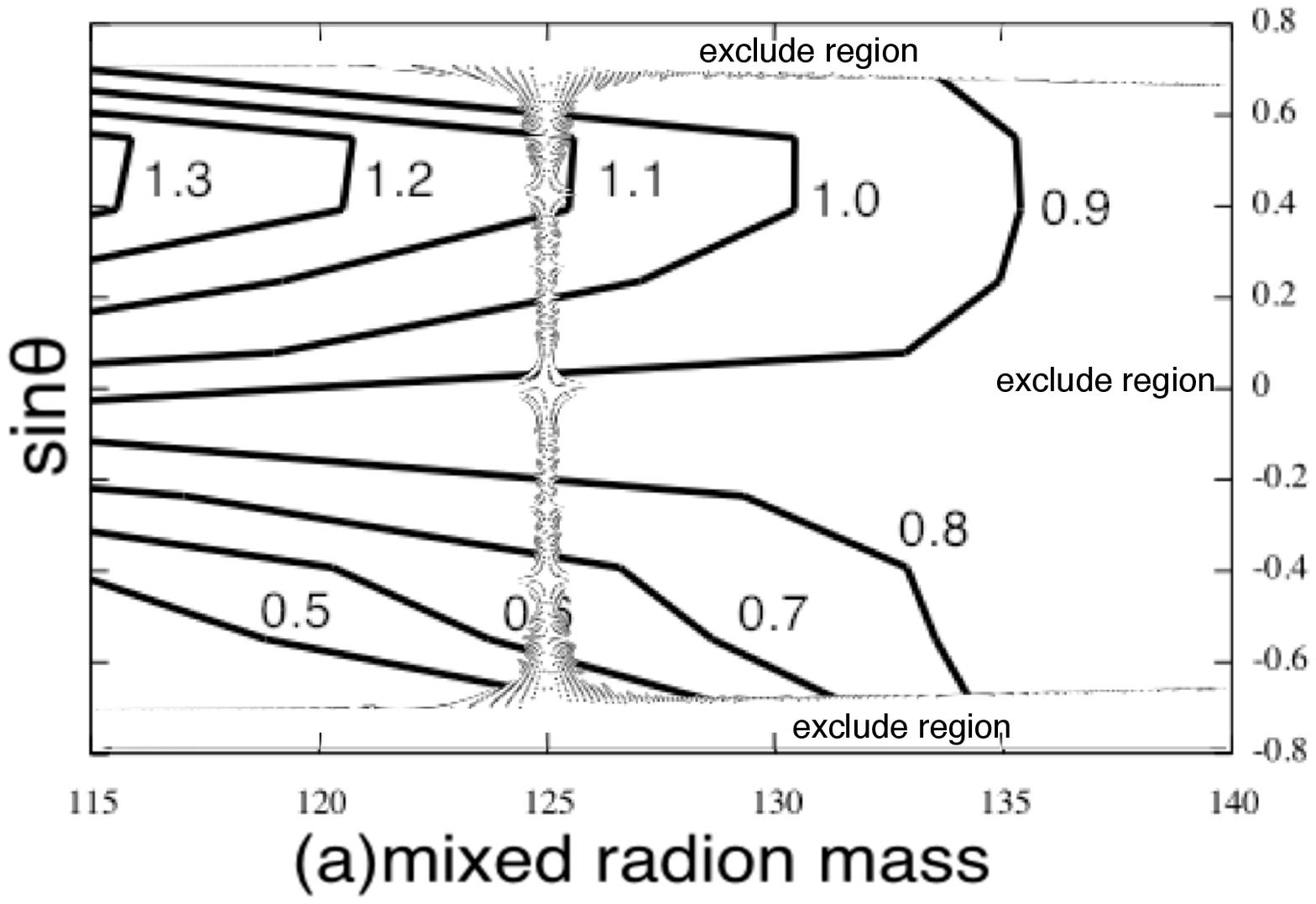}
  \end{center}
 \end{minipage}
 \begin{minipage}{0.45\hsize}
  \begin{center}
   \includegraphics[width=70mm]{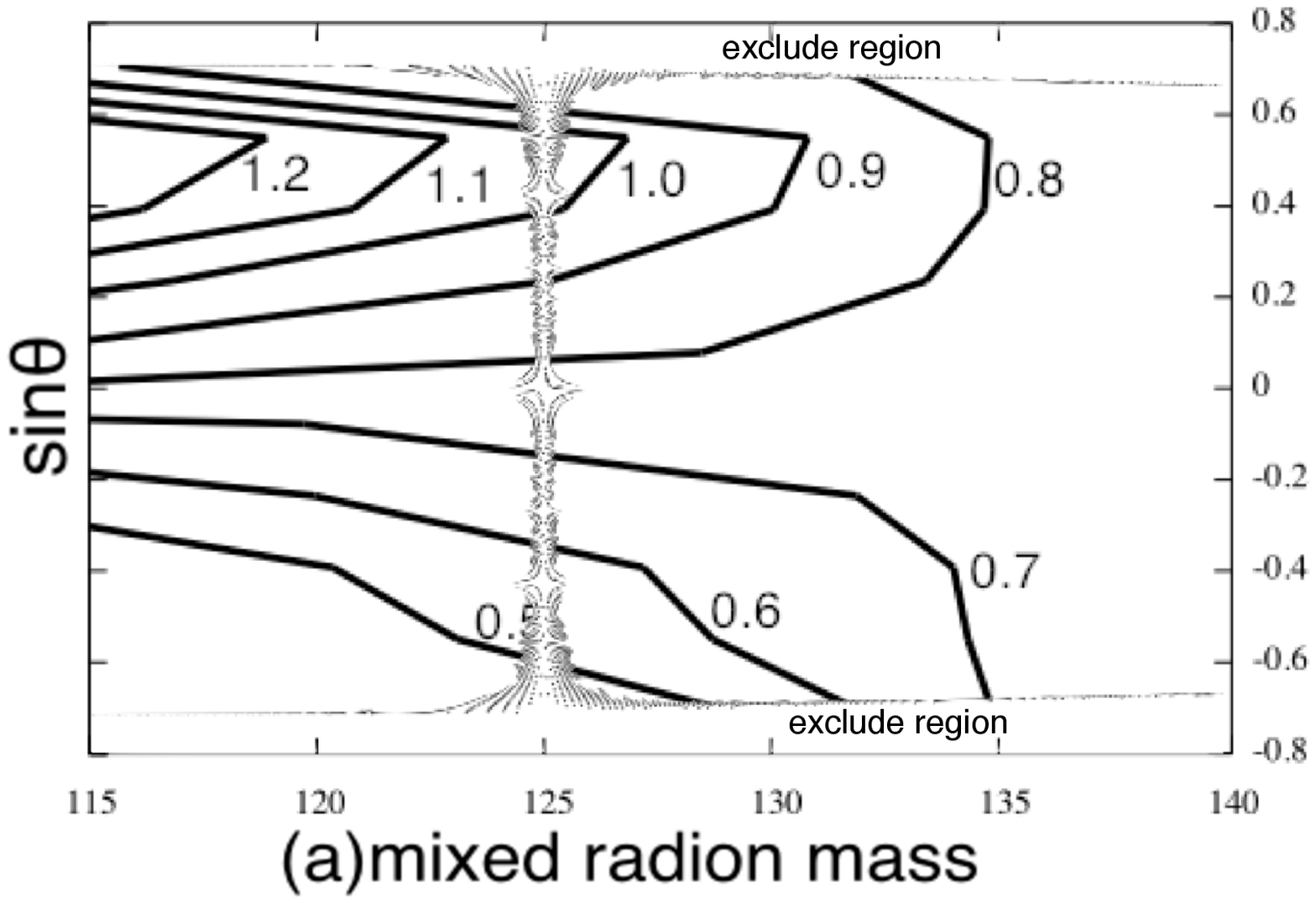}
  \end{center}
 \end{minipage}
 \caption
 {¡¡The Detection ratio of the visible higgs boson for Scenario I. 
 (a) $DR_{h_m}(\gamma \gamma)$ and (b) $DR_{h_m}(ZZ)$ at $\Lambda_\phi=3$ TeV. Both $DR_{h_m}(ZZ)$ and  $DR_{h_m}(\gamma \gamma)$ have the excess in the light $r_m$ (hidden scalar)  and large mixing angle region. }
 \label{fig:four}
\end{figure}

FIG. \ref{fig:four}(a)(b) show (a)$ DR_{h_m}(\gamma \gamma)$ and (b)$DR_{h_m}(ZZ)$ in $m_{r_m}$ and  $\sin\theta$ plane for Scenario I,  $m_{h_{m}}=125$~GeV and $\Lambda_\phi=3$~TeV.
There are excesses of both $DR_{h_m}(ZZ)$ and $DR_{h_m}(\gamma \gamma)$ in the region with light  hidden scalar mass  and large mixing angle. 
In the other region of $m_{r_m}$ and $\sin\theta$,  $DR_{h_m}(ZZ)$ and $DR_{h_m}(\gamma \gamma)$ are suppressed.
Not only $DR$, which is the subject of large QCD uncertainty,
the ratio between $DR_{h_m}(ZZ)$ and $DR_{h_m}(\gamma \gamma)$ is around 1.2 for most of
the parameter region in the figure.
Note that projected sensitivity at 8 TeV 5 fb$^{-1}$ at 95$\%$ CL
for $[\Gamma(gg\rightarrow h)Br(h \rightarrow VV)]/[\Gamma(gg\rightarrow h)Br(h \rightarrow VV)]_{SM}(V=W,Z)$ is well below 0.5 at the LHC.\footnote{PhysicsResultsHIGStandardModelProjections < CMSPublic < TWiki}
The  most of the re3gion given in FIG. \ref{fig:four}(b) would be  accessible in 2012. Sensitivity to $\gamma \gamma$ channel
is less, but
$\Gamma(gg\rightarrow h)Br(h \rightarrow \gamma\gamma)/\Gamma(gg\rightarrow h)Br(h \rightarrow \gamma\gamma)_{SM}>$ 0.8 should be accessible as well.
The region near $m_{r_m}$~125 GeV,  the mass difference between $m_{r_m}$ and $m_{h_m}$ is too small 
to be distinguished, and the detection ratio  must be calculated for the sum of the particles. 
However, due to the cancellation among the diagrams, the contribution  of 
the mixed radion is  small. 
 
\begin{figure}[htbp]
 \begin{minipage}{0.45\hsize}
  \begin{center}
   \includegraphics[width=70mm]{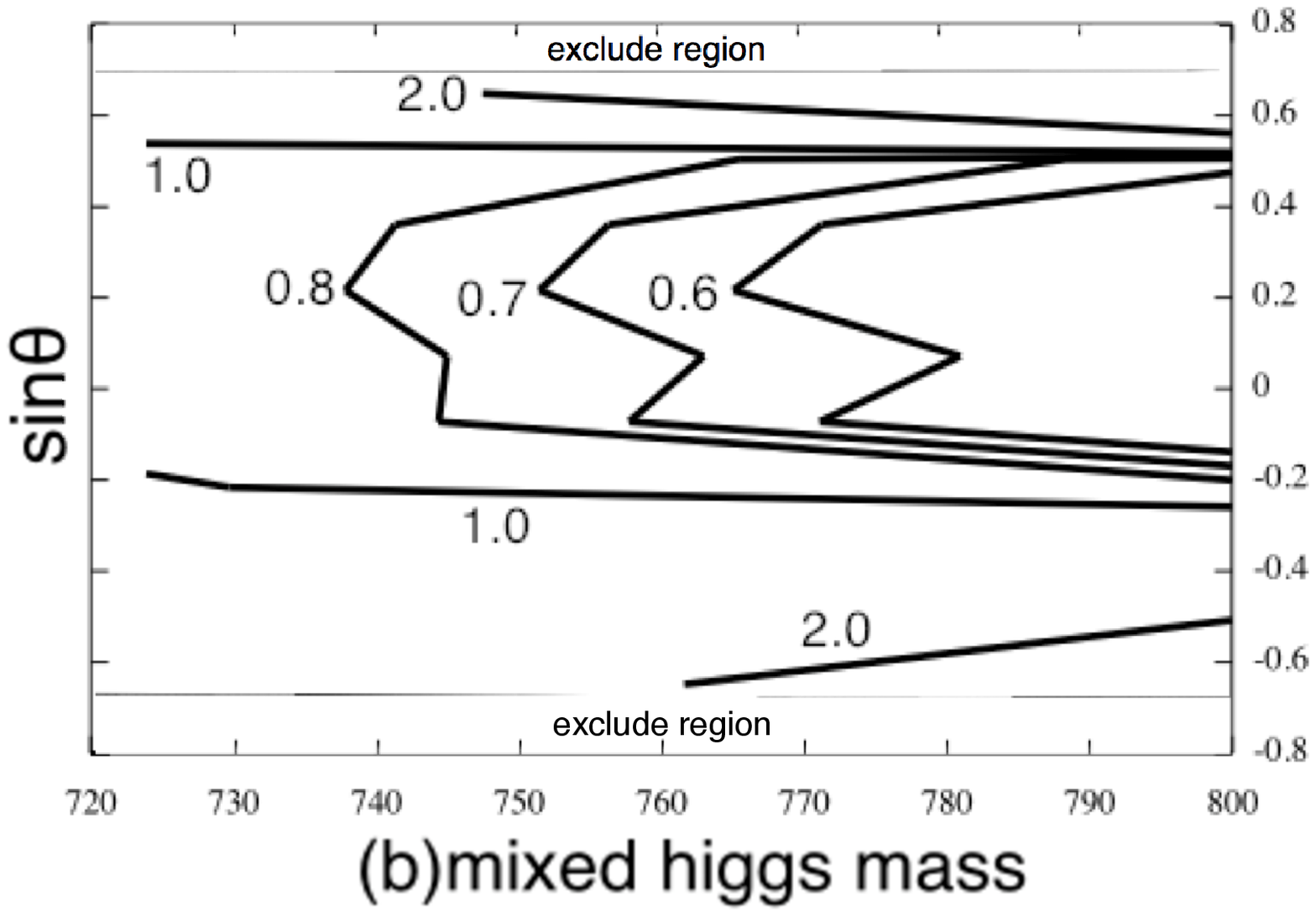}
  \end{center}
 \end{minipage}
 \begin{minipage}{0.45\hsize}
  \begin{center}
   \includegraphics[width=70mm]{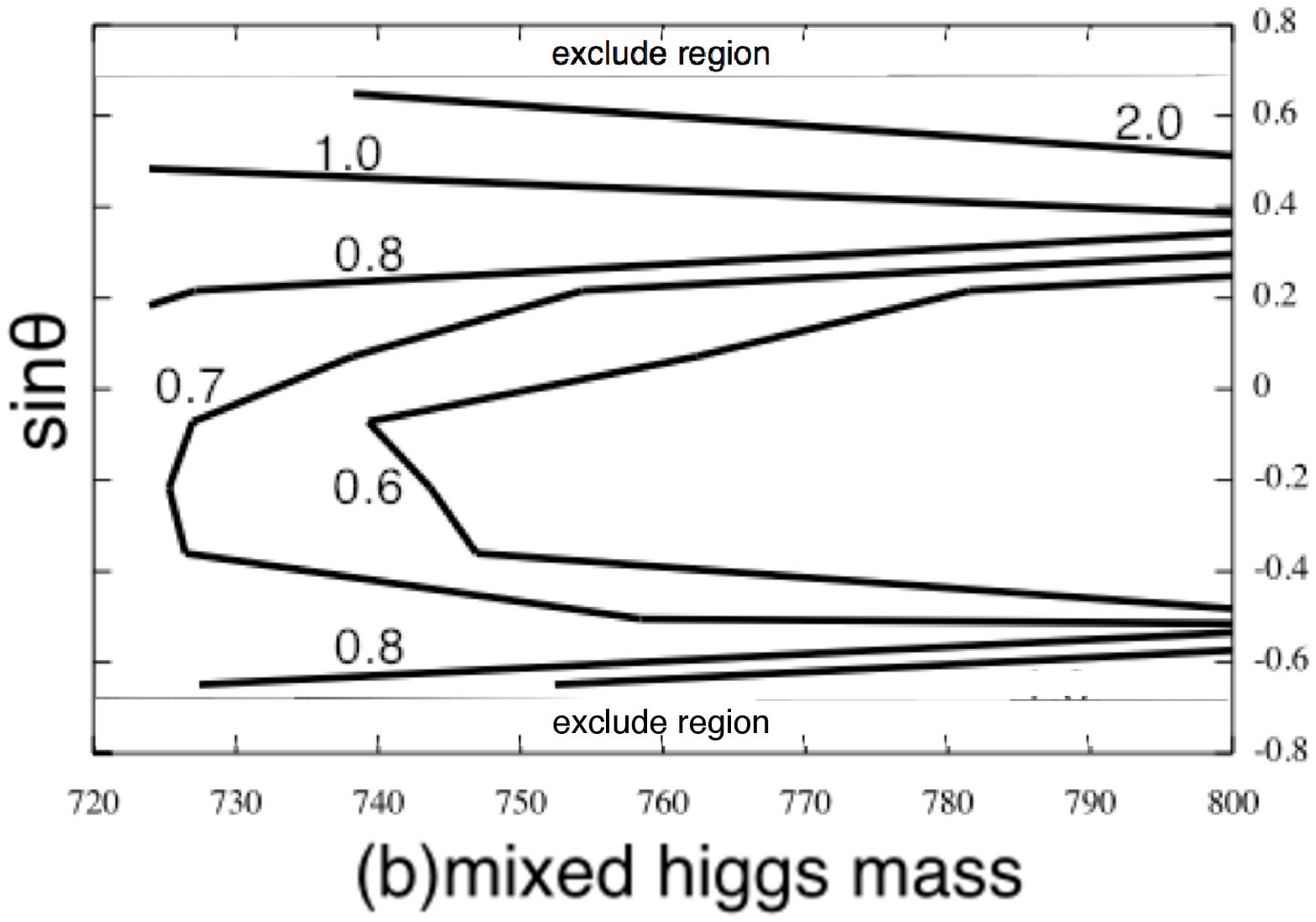}
  \end{center}
 \end{minipage}
 \caption
 { The Detection ratio of the visible higgs boson  for Scenario II. 
  (a) $DR_{r_m}(\gamma \gamma)$ and  (b) $DR_{r_m}(ZZ)$ at $\Lambda_\phi=3$ TeV. Both $DR_{r_m}(ZZ)$ and  $DR_{r_m}(\gamma \gamma)$ have an excess in heavy higgs mass  and large mixing angle region.
 }
 \label{fig:five}
\end{figure}

FIG. \ref{fig:five} shows the corresponding result for Scenario II, a) $DR_{r_m}(ZZ)$ and b) $DR_{r_m}(\gamma \gamma)$ in $m_{h_m}$ and  $\sin\theta$ plane for $m_{r_m}=125$ GeV and $\Lambda_\phi=3$ TeV. There are excesses of both $DR_{r_m}(ZZ)$ and $DR_{r_m}(\gamma \gamma)$ at heavy mixed higgs mass $m_{h_m}>720$ GeV and large mixing angle region $\sin\theta>0.4$.
In the other region of $m_{h_m}$ and $\sin\theta$,  $DR_{r_m}(ZZ)$ and $DR_{r_m}(\gamma \gamma)$ are suppressed.
Here $DR_{r_m}(ZZ)/DR_{r_m}(\gamma \gamma)$ changes drastically, 
because the suppressed region is different in the two modes.   
Excluded regions of FIG. \ref{fig:four} and FIG. \ref{fig:five}  is the same as that of FIG. \ref{fig:two}.  

To show $\Lambda_\phi$ dependence, we calculate the ratio of $BR(h_{\rm vis} \rightarrow \gamma\gamma, ZZ)$ and $\Gamma_{h_{\rm vis}}(gg)$ of our model to those of SM higgs booson as a function of $\Lambda_\phi$. 
FIG. 6(a) shows Scenario I , $\sin\theta=0.5$, $m_{r_m}$=120 GeV and $m_{h_m}=125$~GeV
FIG. 6(b) corresponds to Scenario II, and    we show  the ratio of $BR(r_m \rightarrow \gamma\gamma, ZZ)$ and $\Gamma_{r_m}(gg)$  to those of SM, where 
$\sin\theta=0.5$, $m_{h_m}$=800 GeV. For both cases, 
 $\Gamma_{h_{\rm vis }}(gg)$ is  larger than those of the SM and are suppressed as increasing $\Lambda_\phi$.
Conversely,  $BR(h_{\rm vis} \rightarrow ZZ)$  is smaller than that of the SM. 
In Scenario I , $BR(h_m \rightarrow \gamma\gamma)$  is larger than that of  
the SM, but, in Scenario II , $BR(r_m \rightarrow \gamma\gamma)$ is smaller. 
Therefore, in Scenario I , both large $\Gamma_{h_m}(gg)$ and $BR(h_m \rightarrow \gamma\gamma)$ increase the process $gg \rightarrow h_m \rightarrow ZZ,\gamma\gamma$ and small $BR(h_m \rightarrow ZZ)$ decreases the process $gg \rightarrow h_m \rightarrow ZZ$. 
In Scenario II , $\Gamma_{r_m}(gg)$ is large compared with $\Gamma_h(gg)_{SM}$. 
But, $BR(r_m \rightarrow \gamma\gamma)$ and $BR(r_m \rightarrow ZZ)$ are small.

\begin{figure}[htbp]
 \begin{minipage}{0.45\hsize}
  \begin{center}
   \includegraphics[width=65mm]{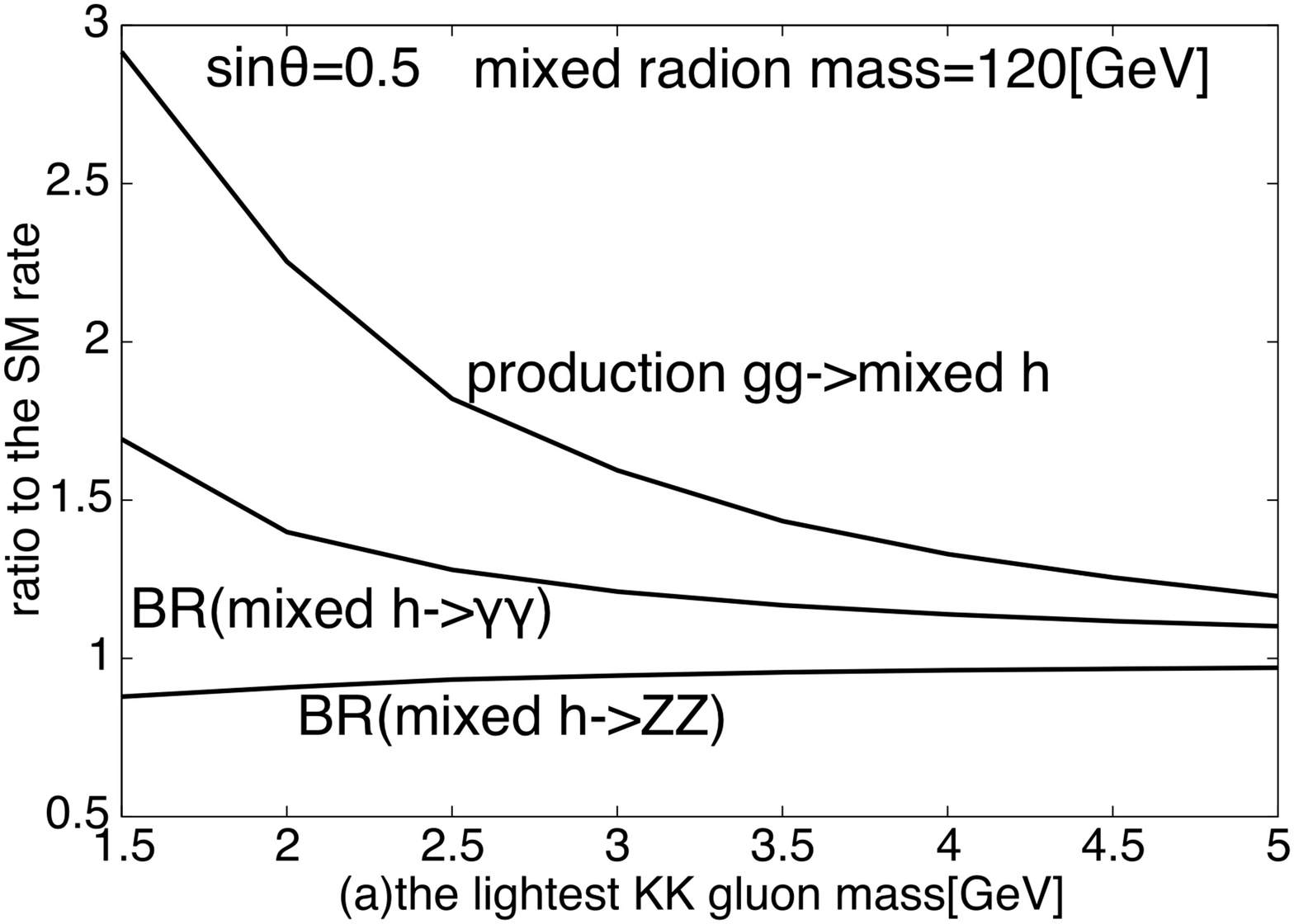}
  \end{center}
 \end{minipage}
 \begin{minipage}{0.45\hsize}
  \begin{center}
   \includegraphics[width=65mm]{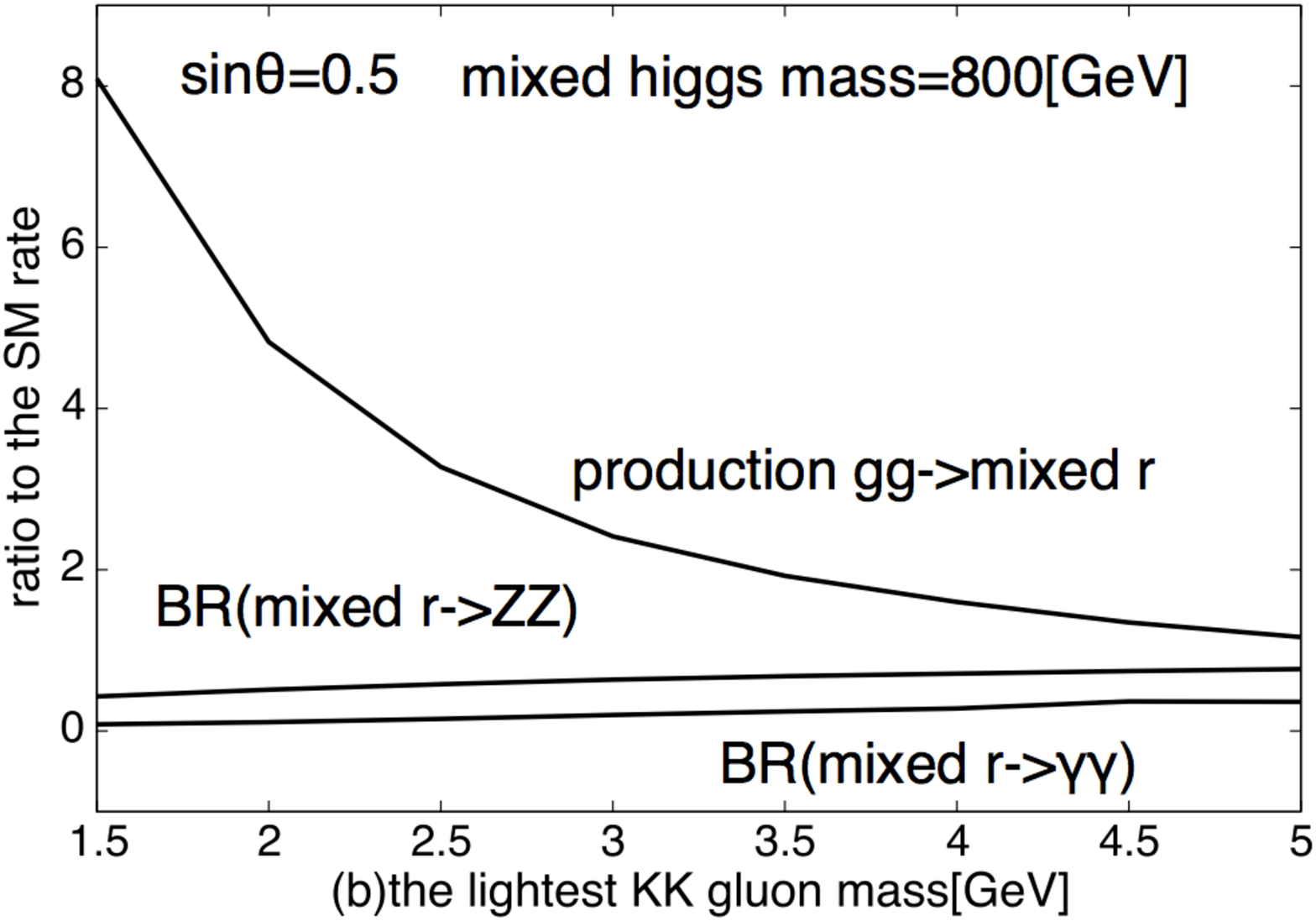}
  \end{center}
 \end{minipage}
 \caption
 {
 ratio of $\Gamma(gg \rightarrow h_m)$, $Br(h_m\rightarrow \gamma \gamma,ZZ)$ to that  of the SM at parameter of excess point in the first scenario (a)($\sin\theta=0.5, m_{r_m}=120$ GeV and $m_{h_m}$=125 GeV). 
The ratio of $\Gamma(gg \rightarrow h_m)$, $Br(h_m\rightarrow \gamma \gamma,ZZ)$ to that  of the SM in the second scenario (b)($\sin\theta=0.5, m_{h_m}=800$ GeV and $m_{r_m}$=125 GeV).
 }\
 \label{fig:six}
\end{figure}

\section{summary and conclusion}
We have investigated the phenomenology of the RS model with the radion and the higgs boson 
mixing in the view of the constraints from the 2011 data at the LHC.  We take a model where the SM particles except the higgs boson live in the bulk.
In the model KK particles of gauge bosons and fermions are predicted to  be at  TeV scale. 
We take into account the KK loop corrections to the production and decay of the higgs boson 
and the radion. 
In this paper we consider only contribution of the KK top in KK fermions.
Profiles of the five dimensional wavefunction of the fermions are determined by five dimensional
mass parameters $c_{L,R}$, and we choose the values of $c_{L,R}$ which maximize the contribution of the KK top. 
Since the interaction of the radion with the SM particles is similar to that of the higgs boson, we may constrain the radian-higgs mixing scenario using LHC data for higgs searches.  
The LHC data in 2011 suggests  excesses of the $ZZ$ and $\gamma \gamma$ channels with respect to the SM higgs at a mass around 125 GeV though it is still statistically  
limited. 

In the first scenario we fix the mass of higgs-like mixed state $m_{h_m}$ at 125 GeV, and scan over 
the mass of the radian-like state $m_{r_m}$ and the mixing angle($\sin\theta$) which have not been disfavored by 2011 data fixing $\Lambda_\phi$ =3 TeV. 
There are no constraints to the mass of $r_m$ at  $\Lambda_\phi$=3 TeV for relatively small $\sin\theta$.
We find the region where the radion mass is relatively light $m_{r_m} <$180 GeV is interesting because the detection ratio
$DR(h_m)$ between 1.6$\sim$ 0.8 (for $\gamma \gamma$)  1.2$\sim$0.6 (for $ZZ$) is still allowed. 
In the second scenario, we fix $m_{r_m}$ at 125 GeV and 
scan over the mass of higgs-like state and mixing angle. 
In this case, we have allowed region where the mass of the higgs like state below 130~GeV,  $200$~GeV$< m_{h_m}<350$~GeV, and $m_{h_m}>550$~GeV. 
The detection ratio of the radion is mostly  around 0.5 to 0.7 for $m_{r_m}>$550~GeV. However, there is a region of parameter space where the ratio is very close to SM higgs value for $\sin\theta < -0.3$ and $\sin\theta> 0.5$ for $r_m\rightarrow \gamma \gamma$ and ($\sin\theta > 0.4$ for $r_m\rightarrow ZZ$) and  $m_{h_m} >$ 700~GeV. 
In 2012, we expect more than 5 times the data of 2012, and the higgs and radion scenario will be 
further constrained. 

\section*{Acknowledgments}
This work is supported by Grant-in-Aid for Scientific research from the Ministry of
Education, Science, Sports, and Culture (MEXT), Japan (Nos. 22540300,
23104006 for M.M.N.), and also by World Premier International Research Center Initiative (WPI Initiative), MEXT, Japan.

\renewcommand{\refname}{References}

\end{document}